\documentclass[aps,twocolumn,
groupedaddress,nofootinbib,nobalancelastpage,nobibnotes]{revtex4}
\pdfoutput=1
\usepackage{amsmath,amsfonts,amssymb,mathrsfs,graphicx}
\usepackage[squaren]{SIunits}
\usepackage{verbatim}
\usepackage{hyperref}
\usepackage{slashed}
\usepackage[utf8]{inputenc}
\usepackage{float}
\usepackage{bbold}
\usepackage{pifont}
\usepackage{url}
\usepackage[dvipsnames]{xcolor}

\usepackage{afterpage}

\usepackage{array}
\usepackage{slashbox}

\usepackage{rotating}

\newcommand*{\affmark}[1][*]{\textsuperscript{#1}}

\newcommand{\ie}{{\it i.e.}}

\newcommand{\eg}{{\it e.g.}}

\newcommand{\cf}{{\it cf.}}

\newcommand{\eq}{Eq.}

\newcommand{\fig}{Fig.}

\newcommand{\Ref}{Ref.}
\newcommand{\Refs}{Refs.}
\newcommand{\Sec}{Section}

\newcommand{\Tab}{Tab.}

\newcommand{\ba}{\begin{array}}
\newcommand{\ea}{\end{array}}

\newcommand{\equ}[1]{\eq~(\ref{eq:#1})}
\newcommand{\figu}[1]{\fig~\ref{fig:#1}}
\newcommand{\tabl}[1]{\Tab~\ref{tab:#1}}
\newcommand{\bi}{\begin{itemize}}
\newcommand{\ei}{\end{itemize}}

\newcommand{\auxNH}[3]{%
  \raisebox{-3.7mm}[0.0mm][5.0mm]{\makebox[0pt][l]{\ding{#1}}}%
  \raisebox{-0.8mm}[0.0mm][0.0mm]{\makebox[0pt][l]{\ding{#2}}}%
  \raisebox{+3.7mm}[7.5mm][0mm]{\makebox{\ding{#3}}}}

\newcommand{\auxIH}[3]{%
  \raisebox{-3.7mm}[0.0mm][5.0mm]{\makebox[0pt][l]{\ding{#1}}}%
  \raisebox{+0.8mm}[0.0mm][0.0mm]{\makebox[0pt][l]{\ding{#2}}}%
  \raisebox{+3.7mm}[7.5mm][0.0mm]{\makebox{\ding{#3}}}}

\newcommand{\auxSL}[1]{\makebox[0pt][l]{\hspace{#1}\large/}}

\newcommand{\dgN}[1]{%
  \ifnum#1=1\auxNH{182}{183}{184}\else%
  \ifnum#1=2\auxNH{182}{183}{174}\else%
  \ifnum#1=3\auxNH{182}{173}{184}\else%
  \ifnum#1=4\auxNH{172}{183}{184}\else%
  \ifnum#1=5\auxNH{172}{173}{184}\else%
  \ifnum#1=6\auxNH{172}{183}{174}\else%
  \ifnum#1=7\auxNH{182}{173}{174}\else%
  \ifnum#1=8\auxNH{172}{173}{174}\fi\fi\fi\fi\fi\fi\fi\fi}

\newcommand{\dgI}[1]{%
  \ifnum#1=1\auxIH{184}{182}{183}\else%
  \ifnum#1=2\auxIH{184}{182}{173}\else%
  \ifnum#1=3\auxIH{184}{172}{183}\else%
  \ifnum#1=4\auxIH{174}{182}{183}\else%
  \ifnum#1=5\auxIH{174}{172}{183}\else%
  \ifnum#1=6\auxIH{174}{182}{173}\else%
  \ifnum#1=7\auxIH{184}{172}{173}\else%
  \ifnum#1=8\auxIH{174}{172}{173}\fi\fi\fi\fi\fi\fi\fi\fi}

\newcommand{\mdN}[1]{%
  \ifnum#1=1{LMH}\else%
  \ifnum#1=2{LM\auxSL{1pt}H}\else%
  \ifnum#1=3{L\auxSL{1pt}MH}\else%
  \ifnum#1=4{\auxSL{0pt}LMH}\else%
  \ifnum#1=5{\auxSL{0pt}L\auxSL{1pt}MH}\else%
  \ifnum#1=6{\auxSL{0pt}LM\auxSL{1pt}H}\else%
  \ifnum#1=7{L\auxSL{1pt}M\auxSL{1pt}H}\else%
  \ifnum#1=8{\auxSL{0pt}L\auxSL{1pt}M\auxSL{1pt}H}\fi\fi\fi\fi\fi\fi\fi\fi}

\newcommand{\mdS}[1]{%
  \ifnum#1=1{123}\else%
  \ifnum#1=2{12\auxSL{-1pt}3}\else%
  \ifnum#1=3{1\auxSL{-0.5pt}23}\else%
  \ifnum#1=4{\auxSL{-0.7pt}123}\else%
  \ifnum#1=5{\auxSL{-0.7pt}1\auxSL{-0.5pt}23}\else%
  \ifnum#1=6{\auxSL{-0.7pt}12\auxSL{-1pt}3}\else%
  \ifnum#1=7{1\auxSL{-0.5pt}2\auxSL{-1pt}3}\else%
  \ifnum#1=8{\auxSL{-0.7pt}1\auxSL{-0.5pt}2\auxSL{-1pt}3}\fi\fi\fi\fi\fi\fi\fi\fi}

\begin{document}

\hspace*{0.85\textwidth} \text{DESY 17-107} 

\title{Astrophysical Neutrinos Flavored with Beyond the Standard Model Physics} 

\author{Rasmus W. Rasmussen\affmark[1]} \thanks{Corresponding author: \href{mailto:rasmus.westphal.rasmussen@desy.de}{rasmus.westphal.rasmussen@desy.de}} 

\author{Lukas Lechner \affmark[2]}

\author{Markus Ackermann\affmark[1]}

\author{Marek Kowalski\affmark[1, 3]}

\author{Walter Winter\affmark[1]}

\affiliation{\affmark[1] DESY, Platanenallee 6, 15738 Zeuthen, Germany \\ \affmark[2] Vienna University of Technology, Department of Physics, Wiedner Hauptstrasse 8-10, A-1040 Wien, Austria \\ \affmark[3] Humboldt-Universit\"at zu Berlin, Institut f\"ur Physik, Newtonstrasse 15, 12489 Berlin, Germany}

\date{\today}

\begin{abstract}
We systematically study the allowed parameter space for the flavor composition of astrophysical neutrinos measured at Earth, including beyond the Standard Model theories at production, during propagation, and at detection. One motivation is to illustrate the discrimination power of the next-generation neutrino telescopes such as IceCube-Gen2. We identify several examples that lead to potential deviations from the standard neutrino mixing expectation such as significant sterile neutrino production at the source, effective operators modifying the neutrino propagation at high energies, dark matter interactions in neutrino propagation, or non-standard interactions in Earth matter. IceCube-Gen2 can exclude about 90\% of the allowed parameter space in these cases, and hence will allow to efficiently test and discriminate models. More detailed information can be obtained from additional observables such as the energy-dependence of the effect, fraction of electron antineutrinos at the Glashow resonance, or number of tau neutrino events. 
\end{abstract}

\maketitle

\section{Introduction}

Physics beyond the Standard Model (BSM) may reveal its nature in extreme environments, at extreme distances, or at extreme energies. Because their Standard Model cross section is low, neutrinos -- compared to other messengers -- can be used to look into the interior of stars and to test propagation effects over cosmological distances. One option to search for BSM physics in the astrophysical neutrino flux is using its flavor composition, which is an experimentally challenging approach, but rather insensitive to astrophysical uncertainties.

The IceCube experiment has discovered high-energy neutrinos from an astrophysical origin \cite{Aartsen:2013bka, Aartsen:2013jdh, Aartsen:2013eka, Aartsen:2014gkd} which opens a new window for multi-messenger astrophysics and testing fundamental properties of neutrinos. Many source candidates may describe the astrophysical neutrinos, see \eg\ \Refs~\cite{Loeb:2006tw,Murase:2013rfa, Winter:2013cla, Razzaque:2013uoa, Ahlers:2013xia, Liu:2013wia, Lunardini:2013gva, Chang:2014hua, Murase:2014tsa, Aartsen:2015rwa, Feyereisen:2016fzb}, even though there is no agreement on their source yet from the experimental perspective~\cite{Aartsen:2014cva, Aartsen:2014ivk, Aartsen:2016oji}. Conceptual insights may be obtained from  energy spectrum and sky distribution~\cite{Lipari:2007su, Laha:2013eev, Kalashev:2014vra, Anchordoqui:2016ewn, Watanabe:2014qua, Chen:2014gxa, Aartsen:2016qcr}. In addition, the neutrino flavor composition may provide information for the production mechanism of the astrophysical neutrinos, as discussed in \Refs~\cite{Xing:2006uk, Pakvasa:2007dc, Choubey:2009jq, Lai:2009ke, Esmaili:2009dz, Xu:2014via, Fu:2014isa, Aartsen:2015ivb, Palladino:2015vna, Palladino:2015zua}, and one can use it to test BSM physics, see \eg\ \Refs~\cite{Barenboim:2003jm, Mena:2014sja, Bustamante:2015waa, Bhattacharya:2010xj, Mehta:2011qb, Shoemaker:2015qul}. A complete analysis of spectrum, neutrino flux normalization and flavor composition is, for example, given in \Refs~\cite{Aartsen:2015knd, Palomares-Ruiz:2015mka, Vincent:2016nut}.

BSM physics can alter the flux and composition of the high-energetic neutrinos, and it has been studied intensely in the literature in many different contexts. Decaying dark matter (DM) may explain the PeV neutrinos seen by IceCube \cite{Esmaili:2013gha, Bhattacharya:2014yha, Cherry:2014xra, Agashe:2014yua, Kopp:2015bfa, Dev:2016qbd, Hiroshima:2017hmy, Borah:2017xgm, Bhattacharya:2017jaw}, and new $\nu$-DM interactions could affect the neutrino flavor composition \cite{deSalas:2016svi, Reynoso:2016hjr, Arguelles:2016zvm, Capozzi:2017auw, Arguelles:2017atb}. Decaying dark matter can also lead to a specific flavor composition due to a resonant behavior in the DM annihilation cross section into neutrinos \cite{Garcia-Cely:2017oco, ElAisati:2017ppn}. Interactions with the cosmic neutrino background \cite{Ioka:2014kca, Ng:2014pca, DiFranzo:2015qea} or non-standard interactions \cite{Blennow:2009rp, Gonzalez-Garcia:2016gpq, Salvado:2016uqu, Marfatia:2015hva} may change the composition, and sterile neutrinos may affect it as well, whether they have eV mass \cite{Hollander:2013im, Brdar:2016thq} or similar mass as the active neutrinos (pseudo-Dirac limit) \cite{Beacom:2003eu, Keranen:2003xd, Esmaili:2009fk, Esmaili:2012ac}. Lorentz and CPT violation are other BSM effects that can affect the neutrino flavor \cite{Colladay:1998fq, Kostelecky:2003xn, Hooper:2005jp, Bustamante:2010nq, Lai:2017bbl, Borriello:2013ala, Stecker:2014xja, Tomar:2015fha, Wei:2016ygk, Wang:2016lne, Ando:2009ts, Arguelles:2015dca}, just as neutrino decays \cite{Beacom:2002vi, Pagliaroli:2015rca, Maltoni:2008jr, Baerwald:2012kc, Bustamante:2016ciw}. Quantum decoherence may be a remnant of a high energy theory such as quantum gravity which might impact the flavor composition \cite{Anchordoqui:2005gj, Morgan:2004vv, Hooper:2004xr}. A more exotic, but plausible phenomenon may be that sterile neutrinos can travel off the Standard Model brane into  extra dimensions, which can change the flavor composition \cite{Aeikens:2014yga, Esmaili:2014esa, 0954-3899-40-5-055202, 0954-3899-40-7-079501, Kisselev2010, Lykken:2007kp}. Note that an unequal neutrino-antineutrino composition can fake new physics \cite{Nunokawa:2016pop} if it is not properly accounted for in the analysis. Furthermore, while these effects are not BSM physics, matter and coherence effects can also alter the flavor composition \cite{Farzan:2008eg, Sahu:2010ap}. Addtional BSM physics can be constrained by IceCube, whether it alters the neutrino flavor composition or not. High-energy neutrinos can resonantly produce TeV-scale squarks, and upper limits on the $R$-parity violating couplings can be derived as a function of the squark's mass \cite{Dev:2016uxj}.  

In this paper we study BSM scenarios and discuss their imprint on the neutrino flavor composition in a systematic matter. We will demonstrate that these scenarios can lead to large deviations from the flavor compositions allowed by standard mixing, and neutrino telescopes can constrain them by measuring the flavor composition. This study is organized as follows: In \Sec~\ref{sec:methods}, we describe our method to systematically address the neutrino flavor composition parameter space. Thereafter, we investigate the imprint of BSM physics and matter effects on the neutrino flavor composition, categorizing the scenarios into three different sections, namely source effects (\Sec~\ref{sec:sourceeffects}), propagation effects (\Sec~\ref{sec:propa}) and detection effects (\Sec~\ref{sec:detectioneffects}). After that, we discuss the exclusion power of neutrino telescopes when investigating the neutrino flavor composition (\Sec~\ref{sec:flavordiscrimination}), energy-dependent effects (\Sec~\ref{sec:energydependence}), fraction of electron antineutrinos at the Glashow resonance (\Sec~\ref{sec:glashowresonance}) or expected number of tau neutrino events (\Sec~\ref{sec:tauevent}). Finally, we summarize in \Sec~\ref{sec:summary}.

\section{Methods}
\label{sec:methods}

One of our main motivations is to study the impact of the next generation neutrino telescopes such as IceCube-Gen2, on the theoretically allowed parameter space expected from physics beyond the Standard Model. Compared to earlier studies, we do not produce figures showing where in parameter space the models for certain sets of parameters lie (``dot plots''), but we show the envelope describing the whole parameter space which is, in principle, allowed -- without weighting certain regions to be more or less likely. 

Given a certain theory beyond the Standard Model, there are typically several uncertainties to describe the allowed parameter space at Earth:
\begin{enumerate}
 \item 
  Each oscillation parameter is only known to a certain precision.
 \item
  The flavor composition at the source is unknown.
 \item
  There is some freedom in the choice of the theory parameters. 
\end{enumerate}
Since we aim to compare the allowed theory parameter space to the expected precision from future experiments, we have to define a coherent approach to address these uncertainties:

\subsection*{Oscillation parameter uncertainties}

In the theory of neutrino oscillation, the flavor composition of neutrinos reaching Earth is given by  
\begin{equation}
 \xi_{\beta+\bar{\beta}, \oplus}=\sum_{\alpha} P_{\alpha \beta} \xi_{\alpha+\bar{\alpha}}
\end{equation}
where $\xi_{\beta+\bar{\beta}, \oplus}=\xi_{\beta, \oplus}+\xi_{\bar{\beta}, \oplus}$ is the final (neutrino+antineutrino) flavor composition and $\xi_{\alpha+\bar{\alpha}}$ is the initial (neutrino+antineutrino) flavor composition. The initial (final) flavor composition of neutrinos and antineutrinos are given by $\xi_{\alpha}$ $(\xi_{\beta, \oplus})$ and $\xi_{\bar{\alpha}}$ $(\xi_{\bar{\beta}, \oplus})$, respectively. Averaged neutrino oscillations lead to flavor mixing (see \Ref~\cite{Farzan:2008eg} for a detailed discussion) such that the transition probability $\nu_\alpha \rightarrow \nu_\beta$ is given by  $P_{\alpha \beta}=\sum_{i=1}^{3}|(U_{\text{PMNS}})_{\alpha i}|^{2}|(U_{\text{PMNS}})_{\beta i}|^{2}$ with $U_{\text{PMNS}}$ being the PMNS neutrino mixing matrix. Therefore, the flavor mixing will only depend on three mixing angles $\theta_{12}, \theta_{23}, \theta_{13}$ and the CP-violating phase $\delta$. 

These oscillation parameters carry uncertainties which translate into uncertainties of the flavor composition at Earth. In order to quantify the present uncertainties,  we define
\begin{equation}
 \chi^{2}= \sum\limits_{j>i} \left( \frac{\text{sin}^{2}\theta_{ij}-\text{sin}^{2}\theta_{ij}^{\text{bf}}}{\sigma_{\text{sin}^{2}\theta_{ij}}}\right)^{2} \, ,
\label{eq:chisquare}
\end{equation}
where the best-fit and uncertainties of the three mixing angles $\theta_{12}$, $\theta_{13}$ and $\theta_{23}$  are taken from \Ref~\cite{Esteban:2016qun}, and we allow for arbitrary values of the CP-violating phase $\delta$. We require $\chi^{2} \leq 11.83$ ($99 \%$ confidence level (CL) for two-dimensional fit), else we disregard the set of oscillation parameters. 

While we use this method to compare between the currently allowed theory parameter space and the IceCube flavor measurement~\cite{Aartsen:2015knd}, the volume upgrade IceCube-Gen2~\cite{Aartsen:2014njl} is a possible future extension with better detector capabilities. The expected sensitivity of IceCube-Gen2 is obtained from \Ref~\cite{MarekGen2} and assumes 15 years of data taking. As the neutrino oscillation parameters will be  known to higher precision at that time, we have to define a corresponding ``Gen2 scenario''. We extrapolate that 
\begin{align}
 \text{sin}^{2}\theta_{12} & =0.306 \pm 0.002 \, , \nonumber \\
 \text{sin}^{2}\theta_{23} &= 0.441 \pm 0.01 \, , \nonumber \\
\text{sin}^{2}\theta_{13}&=0.0217 \pm 0.0005 \, , \nonumber \\ 
\delta &=261^\circ \pm 15^\circ \, , 
\label{eq:benchmark2030}
\end{align}
where the best-fit values are taken from \Ref~\cite{Esteban:2016qun}. The uncertainties are obtained in the following way: The Deep Underground Neutrino Experiment (DUNE) will constrain $\theta_{23}$ and $\delta$ to $\sigma_{\theta_{23}}^{\text{DUNE}} \simeq 1^\circ$ and $\delta^{\text{DUNE}} \simeq 15^\circ$, respectively, by 2027--2028 \cite{Acciarri:2016crz}. The Jianmen Underground Neutrino Observatory (JUNO) will constrain $\sin^{2}(\theta_{12})$ to $\sigma_{\sin^{2}\theta_{12}}^{\text{JUNO}} \simeq 0.003$ by about 2026 \cite{Djurcic:2015vqa, An:2015jdp}. 
We extrapolate these uncertainties to the year 2030 to assure a common level for the neutrino oscillation parameters, by assuming that they scale $\propto 1/\sqrt{\text{exposure}}$. The uncertainty of the reactor angle $(\theta_{13})$ is the most difficult to determine. We assume that the best result will still come from short baseline (SBL) reactor experiments, and the current best measurement is $\sin^{2}(2\theta_{13})=0.0841\pm0.0027\text{(stat)}\pm0.0019\text{(syst)}$ from the Daya Bay experiment \cite{An:2016ses}. Assuming the systematic uncertainty will dominate in the end, a conservative estimate $\sigma_{\sin^{2}(2\theta_{13})}^{\text{SBL}}=0.0019$ is obtained (if that uncertainty cannot be substantially improved). Note that the $\chi^{2}$  for ``Gen2 scenario'' will depend on $\delta$ in addition to $\theta_{12}, \theta_{13}$ and $\theta_{23}$. 

In all cases, we show the results for the normal mass ordering. In the Standard Model, small changes are expected for the inverted ordering because the best-fit value of $\theta_{23}$ changes, see \eg\  \Ref~\cite{Bustamante:2015waa}.
For most of the models discussed, the allowed parameter space for the inverted ordering is identical -- which we checked numerically. However, there are two exceptions, namely neutrino decay and pseudo-Dirac neutrinos: The neutrino flavor composition parameter space changes because the best-fit value of $\theta_{23}$ is different for the inverted ordering.

\subsection*{Unknown flavor composition at source}

For the discussion in this work, it is essential to define the allowed range for the flavor composition at the source within standard mixing. While neutrino production by the pion decay chain leads to the well-known $(\xi_{e+\bar{e}}:\xi_{\mu+\bar{\mu}}:\xi_{\tau+\bar{\tau}})=(1/3:2/3:0)$, the muon decay contribution may be damped by magnetic field effects on the secondaries $(\xi_{e+\bar{e}}:\xi_{\mu+\bar{\mu}}:\xi_{\tau+\bar{\tau}})=(0:1:0)$~\cite{Kashti:2005qa} or enhanced $(1/2:1/2:0)$ to a muon pile-up~\cite{Hummer:2010ai}. Other frequently used assumptions include neutrino production by neutron decay $(1:0:0)$ \cite{Anchordoqui:2003vc} or charmed meson decays $(1/2:1/2:0)$~\cite{Kachelriess:2006fi} at the highest energies. 

In the Standard Model, no significant contribution of tau neutrinos is expected at the source~\cite{Olive:2016xmw}. The main reason is the relatively large mass of the tau lepton, which is the primary of the neutrino. We therefore assume that the initial flavor composition is $(\xi_{e+\bar{e}}:\xi_{\mu+\bar{\mu}}:\xi_{\tau+\bar{\tau}})=(x:1-x:0)$ with $0\leq x \leq1$, unless stated otherwise, whereas a different flavor composition points to physics beyond the Standard Model. This means that we will allow $x$ to vary in that range to describe the allowed region. 

\subsection*{Theory model parameters}

A theoretical model typical comes with unknown theory parameters or choices, which can be either continuous or discrete. Take, for example, neutrino decay. There are $2^3=8$ discrete possibilities, since each mass eigenstate can be stable or unstable, leading to eight different scenarios. There are additional continuous parameters such as the lifetimes of the individual mass eigenstates and the branching ratios into the daughter states, which will lead, over astrophysical distances, to different (continuous) occupations of the daughters. 

We will deal with these discrete and continuous parameters in different ways. In the individual model sections we will show the impact of different discrete choices, while we will vary the continuous parameters in the allowed ranges. In several cases, we will also show the ``complete envelope'' for all possible (discrete and continuous) parameter value choices to later compare different theories with each other, and to show the parameter space which is in principle allowed.

\begin{figure*}[t]
\begin{center}
\begin{tabular}{cc}
\includegraphics[width=0.45\textwidth]{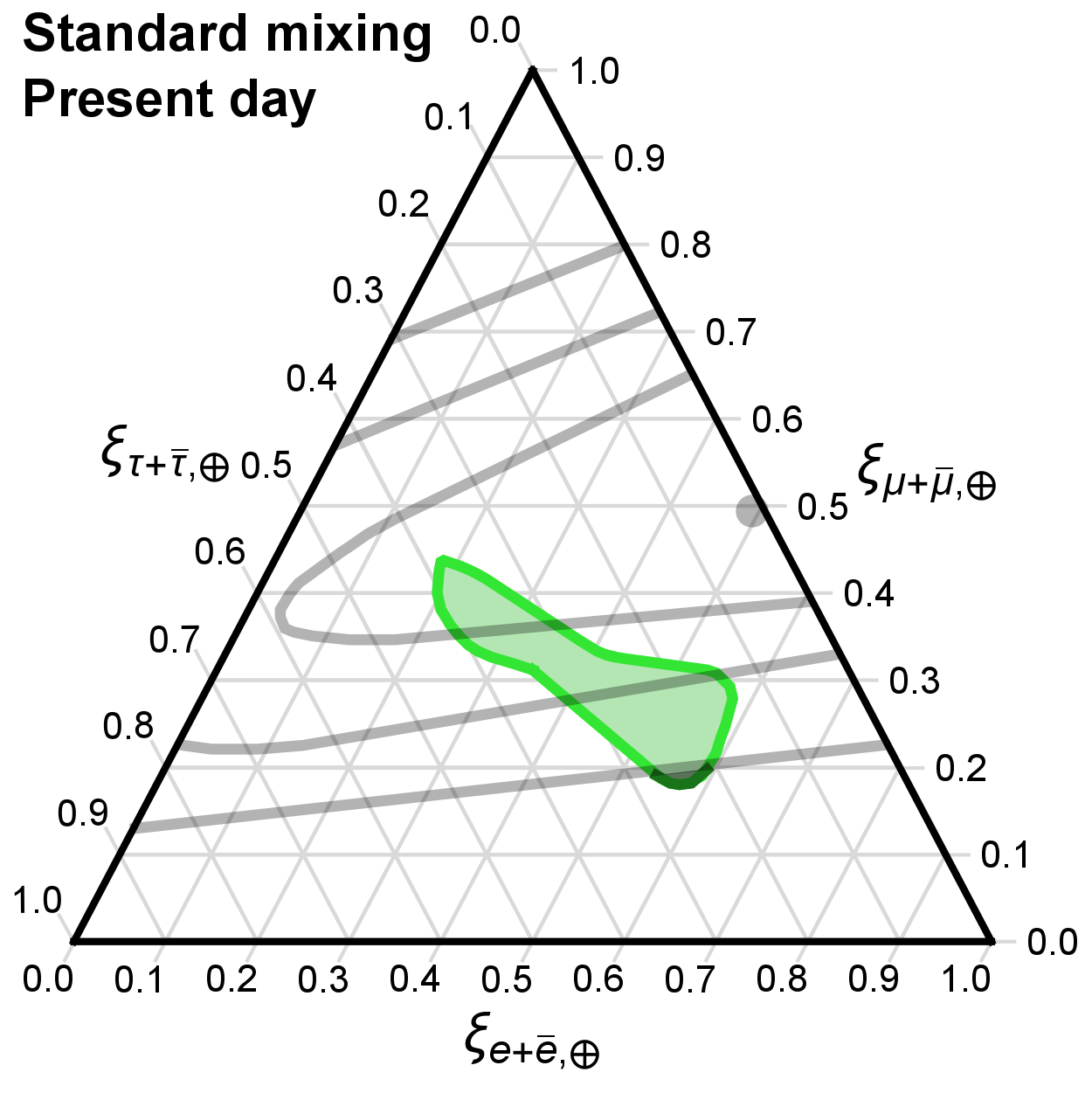} \hspace{0.07\textwidth}
&
\includegraphics[width=0.45\textwidth]{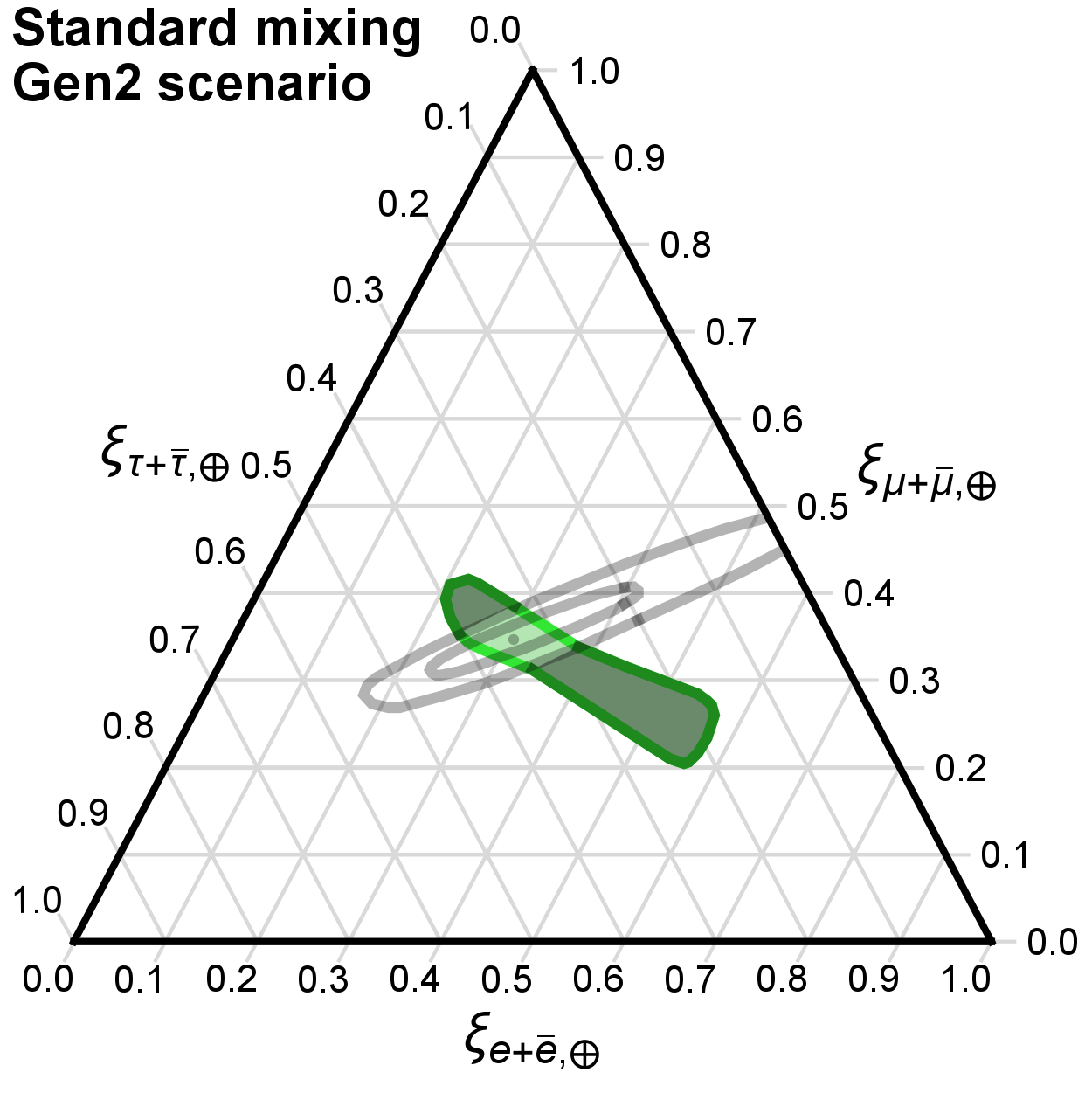}
\\
\end{tabular}
\end{center}
\caption{The flavor composition expected by standard neutrino mixing (green shaded regions) for the current uncertainties (left) and ``Gen2 scenario'' (right) versus the corresponding bounds from IceCube (current) and IceCube-Gen2 (expected), respectively. The different contours corresponds the $1\sigma$, $2 \sigma$ (omitted for IceCube-Gen2), and $3 \sigma$ (2 d.o.f.) allowed regions from \Ref~\cite{Aartsen:2015knd} and \Ref~\cite{MarekGen2}, respectively. The best-fit points are everywhere marked by a dot.}
\label{fig:SM}
\end{figure*}

\subsection*{Graphical representation}

We illustrate our graphical representation for the standard mixing expectation versus current (left panel) and future (right panel) IceCube bounds in \figu{SM}. In this case, the currently published data from the IceCube experiment can exclude 2\% of the standard mixing allowed region at 99\% CL, whereas IceCube-Gen2 will be able to exclude 73\% of the standard mixing allowed region in 2030.\footnote{
One can decompose the flavor triangle into smaller triangles and compute the parameter space area with the IceCube-Gen2 sensitivity region for an increasing number of smaller triangles. This gives an estimate for the exclusion percentage. 
}
The corresponding exclusion regions are marked by darker shadings in \figu{SM}: they correspond to the green shaded regions which our outside the $3\sigma$ allowed contours.

We will in most cases show the ``Gen2 scenario'' only, using the constraints on the oscillation parameters derived above. However, some models, such as those including sterile neutrinos or non-standard neutrino production, have been derived with dedicated models in the literature, making it difficult to extrapolate current uncertaintites. In these cases, which we will explicitly point out, we will use the current uncertainties.

Our discussion in the next sections is separated into source, propagation, and detection effects.

\section{Source effects}
\label{sec:sourceeffects}
Here we discuss effects at neutrino production or close to the source. 

\subsection{Non-standard neutrino production}
\label{sec:NSnuProd}

\begin{figure*}[t]
\centering
\begin{tabular}{c c}
\includegraphics[width=0.32\textwidth]{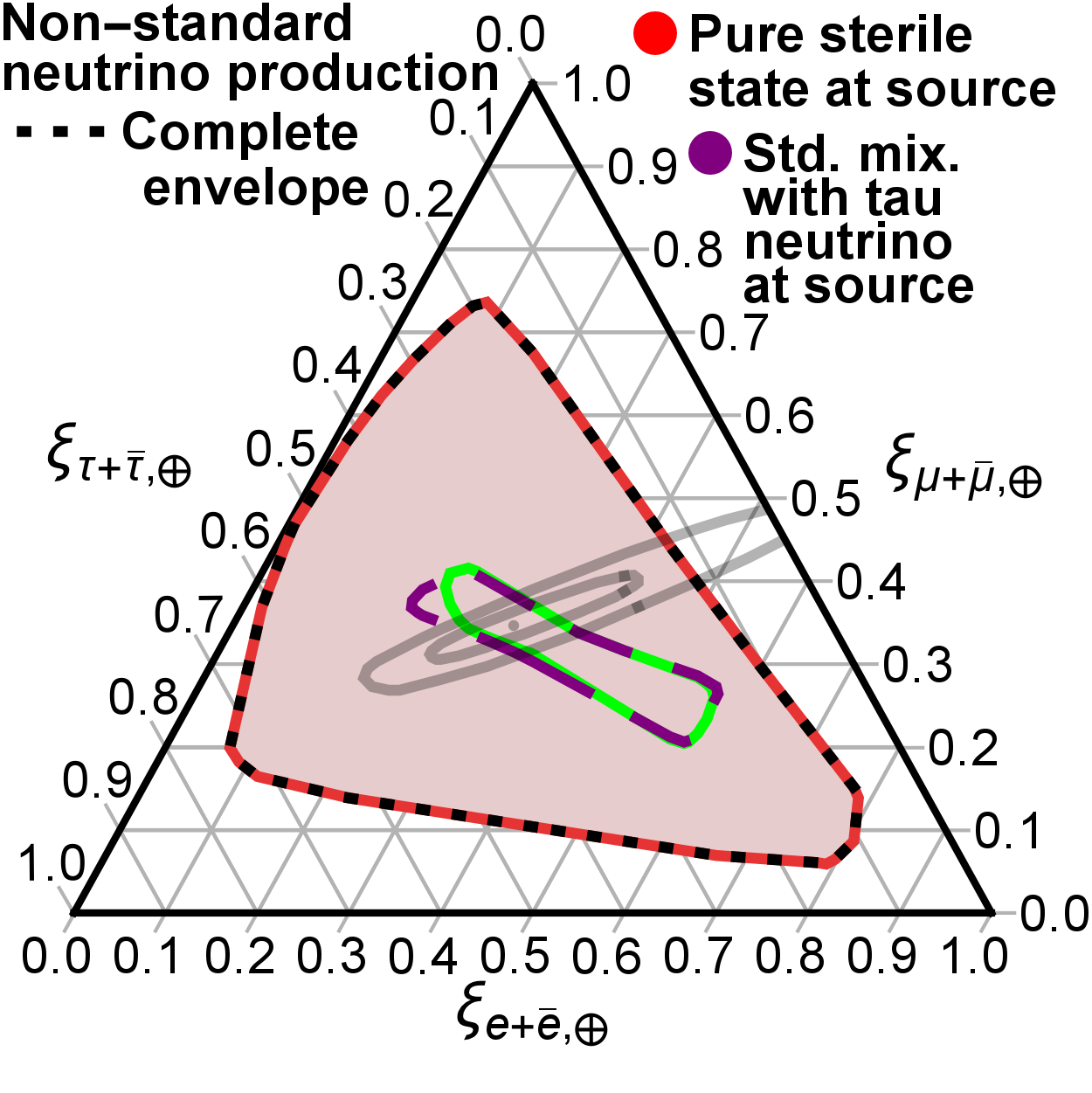}\hspace{0.01\textwidth}
\includegraphics[width=0.32\textwidth]{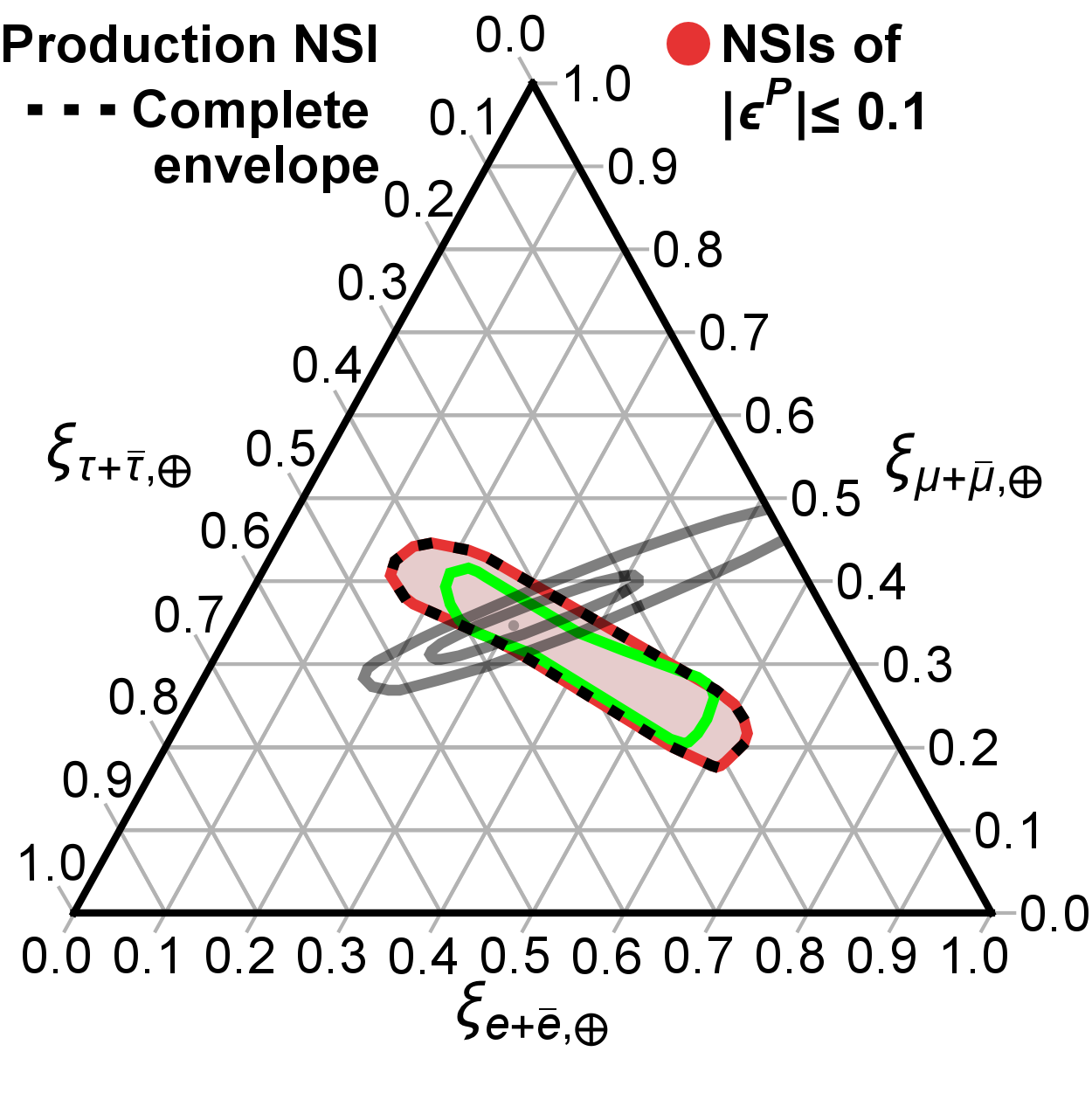}\hspace{0.01\textwidth}
\includegraphics[width=0.32\textwidth]{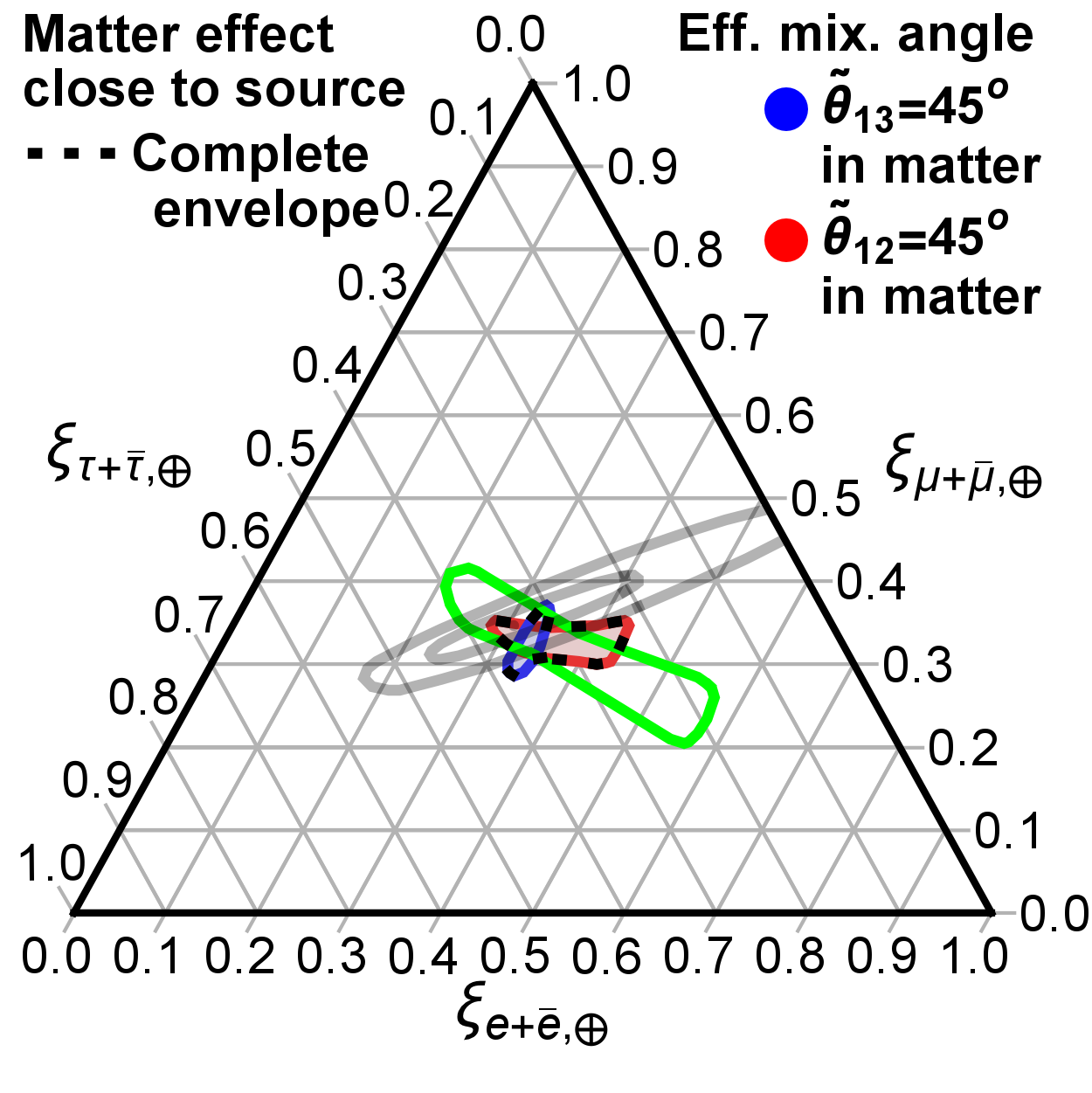}
\\
\end{tabular}
\caption{The neutrino flavor composition at detection for effects at the source: non-standard neutrino production at the source (left), non-standard interactions at production (middle) and matter effect conversions close to the source (right). Green regions mark the standard mixing expectation, gray contours the IceCube-Gen2 expected sensitivity ($1\sigma$, $3\sigma$) for the ``Gen2 scenario''. The best-fit points are everywhere marked by a dot. The dashed contours mark the ``complete envelope'', which is the parameter space in principle allowed -- which is used for reference later.
}
\label{fig:NSprod}
\end{figure*}

One possibility for physics beyond the SM is that it indirectly enters via new (incoherent) production channels which lead to a significant amount of tau neutrinos or sterile neutrinos, which we illustrate in \figu{NSprod}, left panel. 

While tau neutrinos do not affect the standard mixing region very much due to the large $\theta_{23}$ mixing, a substantial amount of sterile neutrinos at production can change the flavor composition~\cite{Brdar:2016thq}. A possible production mechanism may be dark matter which dominantly decays into sterile neutrinos~\cite{Tang:2016sib, Escudero:2016ksa, Escudero:2016tzx}. 
In order to illustrate that, we construct a $4\times4$ mixing matrix using the parameterization
\begin{equation}
 U_{4\times4}=U_{23}\bar{U}_{13}U_{12}\bar{U}_{14}\bar{U}_{24}U_{34} \, ,
\label{eq:extendedmixmatrix} 
\end{equation}
where $U_{ij}$ $(\bar{U}_{ij})$ is a real (complex) rotation matrix in the $ij$ plane. The flavor mixing is $P_{\alpha \beta}=\sum_{i=1}^{4}|(U_{4\times4})_{\alpha i}|^{2}|(U_{4\times4})_{\beta i}|^{2}$ assuming that the sterile neutrino oscillation averages out. One complication is that the $\chi^{2}$ in \equ{chisquare} assumes an unitary $3 \times 3$ mixing matrix to constrain the active mixing angles. Therefore, we construct a new $\chi^{2}$ for this scenario, which is given by 
\begin{equation}
 \chi^{2} = \sum\limits_{\alpha = e, \mu , \tau} \, \sum\limits_{i=1,2,3} \left(\frac{|U_{\alpha i}|-|U_{\alpha i}|^{\text{bf}}}{\sigma_{|U_{\alpha i}|}}\right)^{2} \label{eq:sterilechisquare} 
\end{equation}
where the best-fit values and $1\sigma$ uncertainties are taken from \Ref~\cite{Parke:2015goa}. The authors in \Ref~\cite{Parke:2015goa} investigate the allowed range of the PMNS mixing elements in the case that it is a submatrix of the complete mixing matrix. The mixing matrix can be enlarged due to new physics such as sterile neutrinos, and this will impact neutrino oscillations since the PMNS mixing matrix is non-unitary. Parameterizing the non-unitary neutrino oscillation equation and using the Cauchy-Schwarz inequality to reduce the parameter space, they construct a $\chi^{2}$ depending on the non-unitary neutrino oscillation and use neutrino oscillation data as input to constrain the allowed range of the PMNS mixing matrix elements. Their result is used in our method to obtain the allowed parameter space of the neutrino flavor composition in the cases with a sterile neutrino. Note that we did not extrapolate the uncertainties to 2030 in this case. Our method gives results similar to those shown in \fig~3 of \Ref~\cite{Brdar:2016thq}. 

From \figu{NSprod}, left panel, it can be seen that a substantial amount of sterile neutrinos at the source can lead to strong deviations from the standard mixing region. IceCube-Gen2 can exclude 93\% of the (currently) allowed parameter space at the 99\% CL.

\subsection{Non-standard interactions at production}

\label{sec:NSIProd}
Non-standard interactions (NSIs) can occur at production of the neutrinos. Such non-standard interactions come from effective higher dimensional operators, which means that BSM physics enters the production process directly. Note that neutrino production and detection processes are typically different, which means that it is plausible that BSM physics affects one or the other, or both.  The neutrino state can be written as \cite{Ohlsson:2012kf}
\begin{align}
 | \nu_{\alpha}^{P} \rangle & = (1+\epsilon^{P})U_{\text{PMNS}}|\nu_{i} \rangle, \\
  \langle \nu_{\beta}^{D} | & = \langle \nu_{i} | U_{\text{PMNS}}^{\dagger}
\end{align}
where $|\nu_{\alpha}^{P} \rangle$ ($\langle \nu_{\beta}^{D}|$) represents the neutrino at production (detection), $|\nu_{i} \rangle$ is the mass eigenstate, $U_{\text{PMNS}}$ is the PMNS mixing matrix and $\epsilon^{P}$ are the NSIs at production. After averaging out the mass square differences $\Delta m_{21}^{2}$ and $\Delta m_{32}^{2}$, the flavor mixing is given by \cite{Ohlsson:2012kf}
\begin{equation}
 P_{\alpha \beta}=\sum_{i}|\mathcal{J}_{\alpha \beta}^{i}|^{2}
\end{equation}
with
\begin{align}
 \mathcal{J}_{\alpha \beta}^{i}& =(U_{\text{PMNS}})_{\alpha i}^{*}(U_{\text{PMNS}})_{\beta i}+ \nonumber \\
& + \sum_{\gamma}\epsilon^{P}_{\alpha \gamma}(U_{\text{PMNS}})_{\gamma i}^{*}(U_{\text{PMNS}})_{\beta i}.
\end{align}
We assume complex production NSIs satisfying $|\epsilon^{P}|\leq 0.1$ since it represents the value allowed by current constraints \cite{Biggio:2009kv, Ohlsson:2012kf}, and we quantify the allowed neutrino oscillation parameters by \equ{chisquare}. With this, we compute the flavor composition which is shown in \figu{NSprod} (middle plot). There are small deviations from standard mixing for the value allowed by current experimental data. 

\subsection{Constant matter effects close to source}

We consider neutrino oscillations in matter close to the source as a possible mechanism to change the flavor composition of astrophysical neutrinos. Note that this mechanism is exceptional in this work, because it is strictly speaking physics of standard neutrino oscillations together with a special astrophysical environment close to the source. It is however interesting to discuss it in the context of source effects, which may, for instance, occur for hidden astrophysical jets~\cite{Razzaque:2009kq, Fraija:2015gaa, Fraija:2013cha, Murase:2013ffa, Sahu:2010ap}.

For simplicity, we consider SM matter effects~\cite{Mikheev:1986gs, Wolfenstein:1977ue} in a constant matter density close to the source with a matter potential $V_{e}$ in the Hamiltonian
\begin{equation}
  \mathcal{H}_{\text{tot}}=\frac{1}{2E} U
\begin{pmatrix}
0 & 0 & 0 \\
0 & \Delta m_{21}^{2} & 0 \\
0 & 0 & \Delta m_{32}^{2}
\end{pmatrix}
U^{\dagger}+
\begin{pmatrix}
 V_{e} & 0 & 0 \\
0 & 0 & 0 \\
0 & 0 & 0
\end{pmatrix} \, .
\end{equation}
with $U=U_{\text{PMNS}}$.
The neutrino energy is chosen to be $E=100$~TeV and the mass square differences are chosen from their $3\sigma$ ranges~\cite{Esteban:2016qun}. The flavor mixing is given by 
\begin{align}
 & P_{\alpha \beta}\left(L_{\text{vac}}, L_{\text{m}}\right)  =|\langle \nu_{\beta}|U_{\text{PMNS}} e^{-i\mathcal{H}_{\text{vac}}L_{\text{vac}}} U_{\text{PMNS}}^{\dagger}  \nonumber \\
& \qquad U_{\text{m}} e^{-i\mathcal{H}_{\text{m}}L_{\text{m}}} U_{\text{m}}^{\dagger}| \nu_{\alpha} \rangle |^{2} \, ,
\end{align}
where $\mathcal{H}_{\text{m}}= U_{\text{m}}^{\dagger}\mathcal{H}_{\text{tot}}U_{\text{m}}=\frac{1}{2E}\text{diag}(0, \Delta m_{21, \text{eff}}^{2}, \Delta m_{32, \text{eff}}^{2})$ with ``m'' as subscript for matter, $\mathcal{H}_{\text{vac}}=\frac{1}{2E}\text{diag}(0, \Delta m_{21}^{2}, \Delta m_{32}^{2})$ and $L_{\text{vac}}$ $(L_{\text{m}})$ is the distance in vacuum (matter). We use our procedure given in \Sec~\ref{sec:methods} to constrain the neutrino oscillation parameters. To include various neutrino production positions within the source and the effect of decoherence, the flavor mixing is averaged over the distances
\begin{equation}
 \bar{P}_{\alpha \beta}=\int_{0}^{L_{\text{vac}}}\int_{0}^{L_{\text{m}}}P_{\alpha \beta}\left(L_{\text{vac}}^{\prime}, L_{\text{m}}^{\prime}\right) \frac{dL_{\text{vac}}^{\prime}}{L_{\text{vac}}} \frac{dL_{\text{m}}^{\prime}}{L_{\text{m}}}. \nonumber
\end{equation}
The matter distance $L_{\text{m}}$ can be arbitrary since we do not have any information about the size of the source, however we limit it to the range $L_{\text{m}} \in [0,10^{10}]~\text{km}$. The vacuum distance $L_{\text{vac}}$ has to obey $L_{\text{vac}} \gg L_{\text{coh}} \simeq 2E/\Delta m_{21}^{2}$ to include decoherence, therefore we parameterize $L_{\text{vac}}=[10,100]L_{\text{coh}}$. The lower limit has to be larger than one to ensure decoherence, whereas changing the upper limit does not influence the flavor composition (we checked this numerically). We study two cases, one with the electron matter density being equal to $N_{e}=2 \cdot 10^{18}$~cm$^{-3}$ and the other with $N_{e}=8.7 \cdot 10^{19}$~cm$^{-3}$. The former (later) case gives a matter resonance in the source for the $\theta_{12}$ ($\theta_{13}$) mixing angle for a $100$~TeV neutrino. These electron densities are low compared to the Earth density, $N_{e}^{\text{Earth}} \approx 10^{24}$~cm$^{-3}$, and core density of Gamma Ray Bursts (GRBs) and Supernovae (SNe), $N_{e}^{\text{core}} \approx 10^{33}$~cm$^{-3}$. However, one finds the resonance electron density in GRBs and SNe at radii $r \approx 10^{12}$~cm \cite{Lunardini:2000swa, Mena:2006eq, 1990ApJ, 1991ApJ}. We have also studied the matter resonance for the $\theta_{23}$ mixing angle in this setup, but it has very little impact on the neutrino flavor composition due to $\theta_{23}$ being nearly $45^\circ$ in vacuum, which means that the effective $\theta_{23}$ in matter will also be $45^\circ$.

The flavor composition at the detector for the matter effect is displayed in \figu{NSprod} (right) for the two cases described above. The regions tend to be smaller than the one allowed by standard mixing because the constant matter density was fine-tuned such that the effective mixing angle in matter was $45^{\circ}$. They can, however, slightly leave the standard mixing allowed region. 

A varying matter density may alter the flavor composition, however it requires a matter density resonance and a minimum matter width/distance \cite{Lunardini:2000swa} for matter effects to be important. This is not satisfied in optically thin sources where the neutrino is accompanied by an electromagnetic component. However, it can be for optically thick sources since the matter density is higher, but no electromagnetic counterpart is present \cite{Mena:2006eq}. An example is an astrophysical source with a choked jet, a scenario that has been studied in the context of Gamma Ray Bursts (GRBs) \cite{Meszaros:2001ms, Razzaque:2003uv}, Supernovae (SNe) \cite{Razzaque:2004yv} and Active Galactic Nuclei (AGNs) \cite{AlvarezMuniz:2004uz, Stecker:2005hn}. According to \Ref~\cite{Mena:2006eq}, the shower-to-track ratio changes in optically thick sources, depending on source properties and neutrino oscillation parameters. However, a similar study \Ref~\cite{Razzaque:2009kq} finds a different, but more significant result. The two studys differ due to different treatment of averaging and coherence loss, and \Ref~\cite{Razzaque:2009kq} includes additional effects not considered in \Ref~\cite{Mena:2006eq}. Therefore, varying matter effects may considerably modify the flavor composition, whereas a constant matter density has some effect.  

\section{Propagation effects}
\label{sec:propa}

\begin{figure*}[tp]
\begin{center}
\begin{tabular}{c c}
\includegraphics[width=0.4\textwidth]{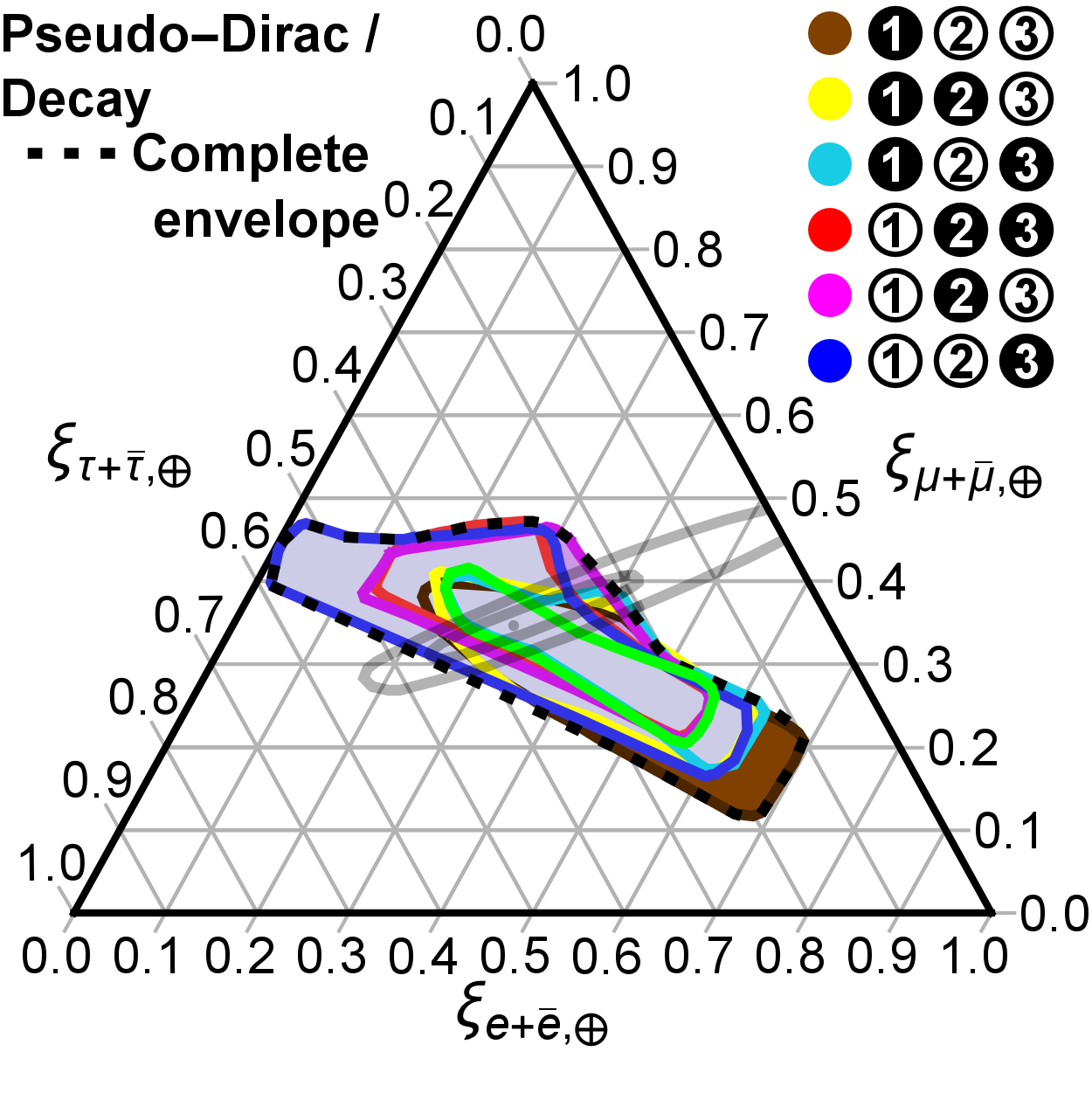}\hspace{0.07\textwidth}
&
\includegraphics[width=0.4\textwidth]{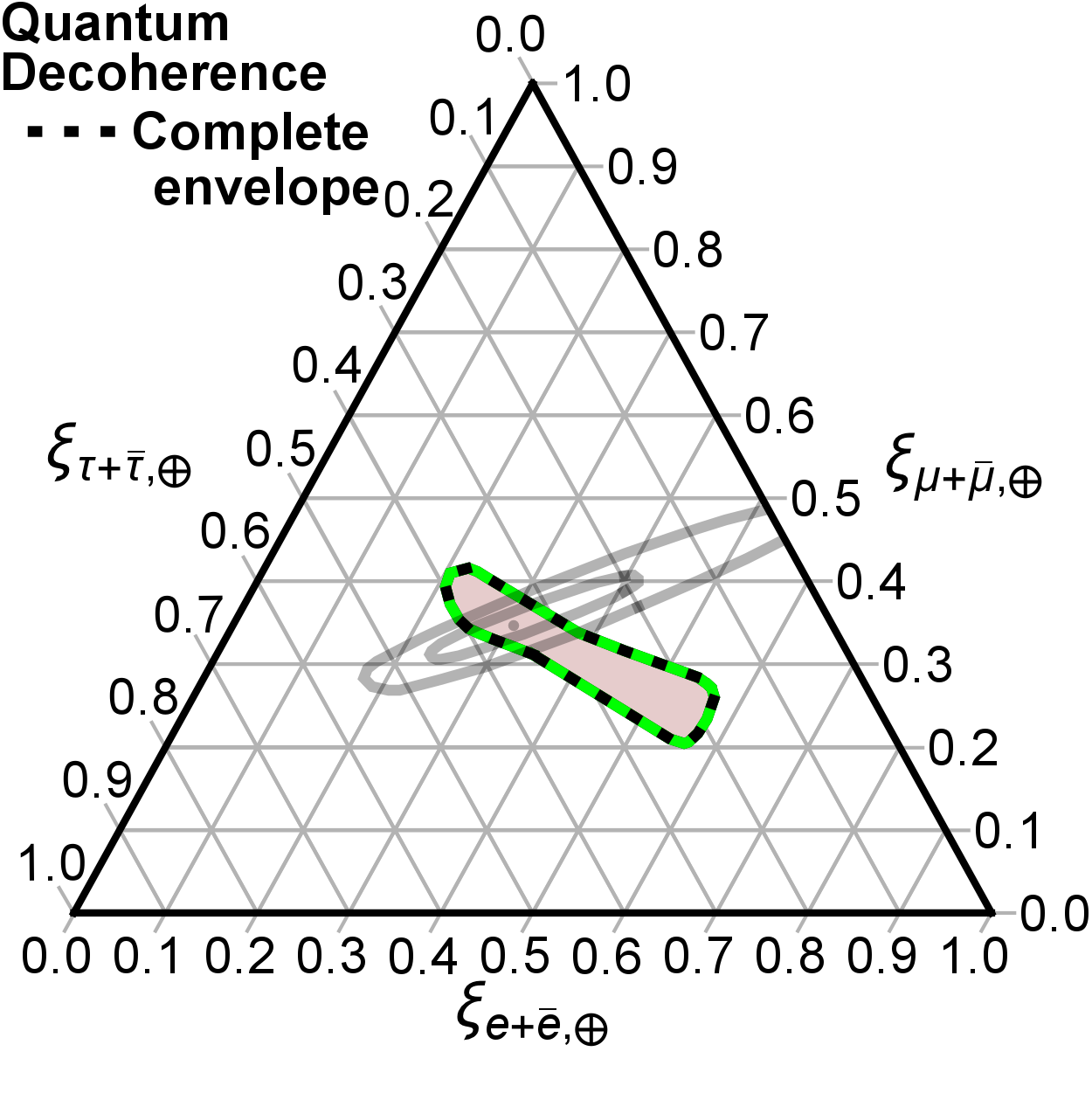}
\\
\includegraphics[width=0.4\textwidth]{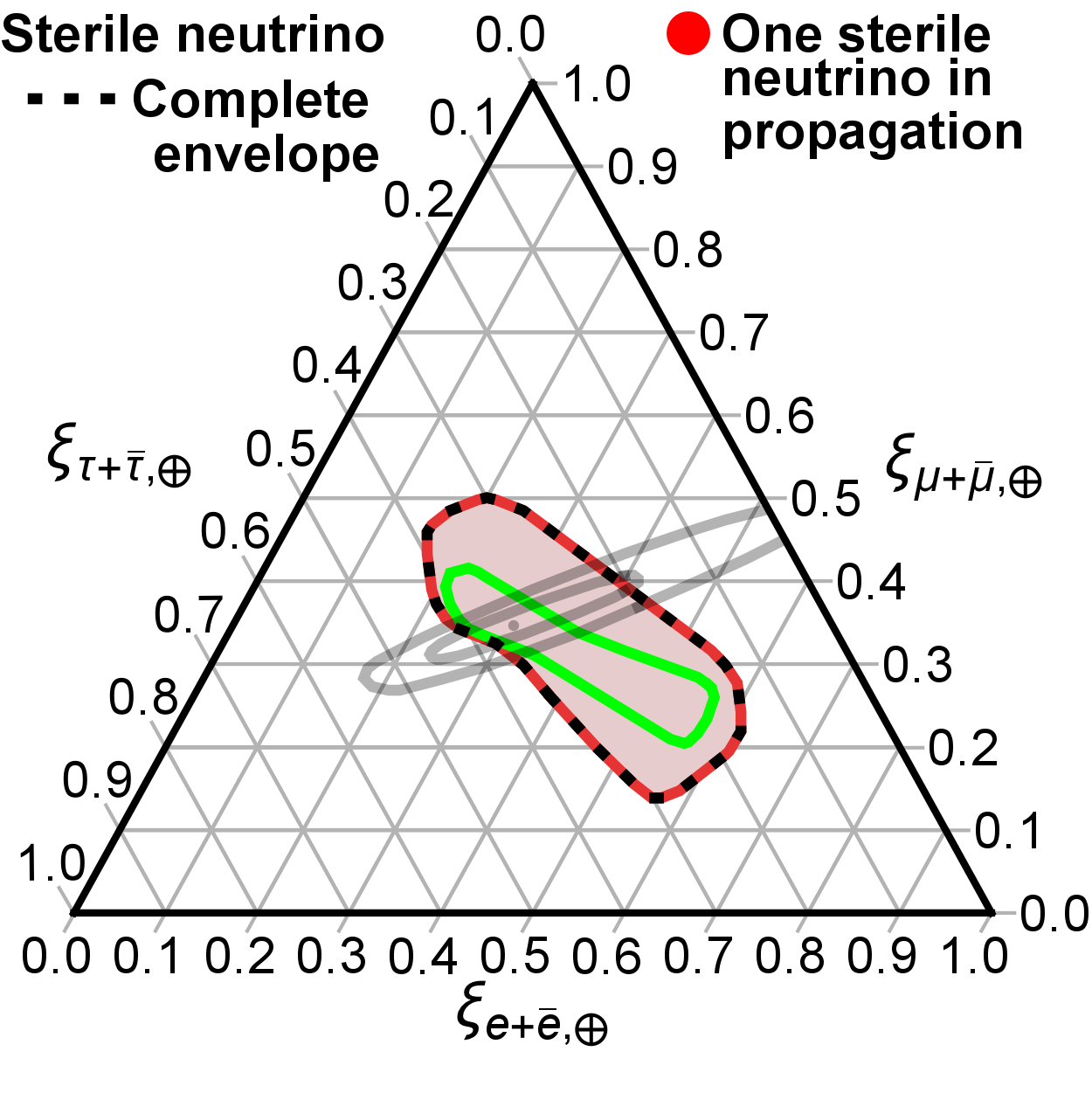}\hspace{0.07\textwidth}
&
\includegraphics[width=0.4\textwidth]{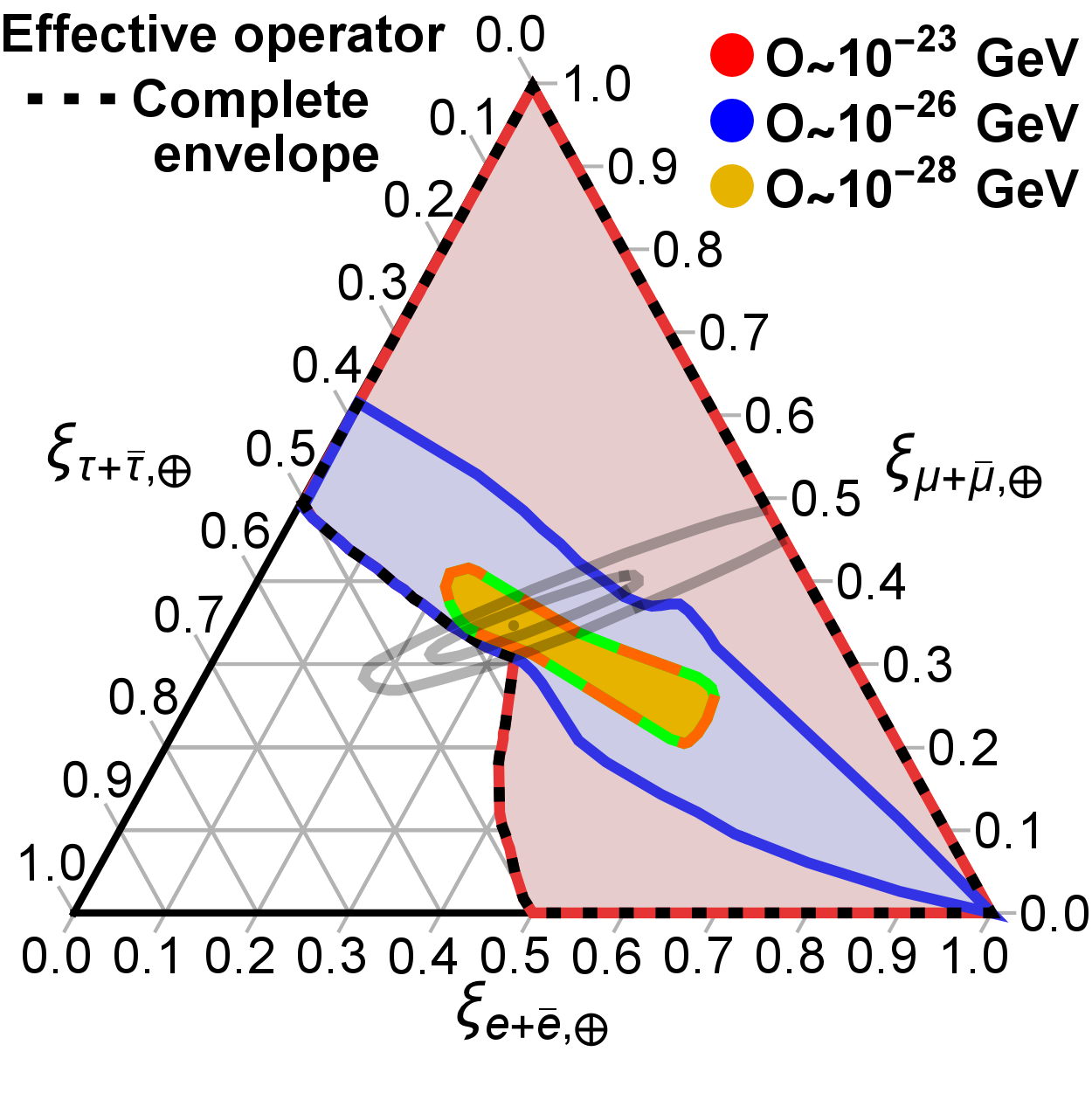}
\\
\includegraphics[width=0.4\textwidth]{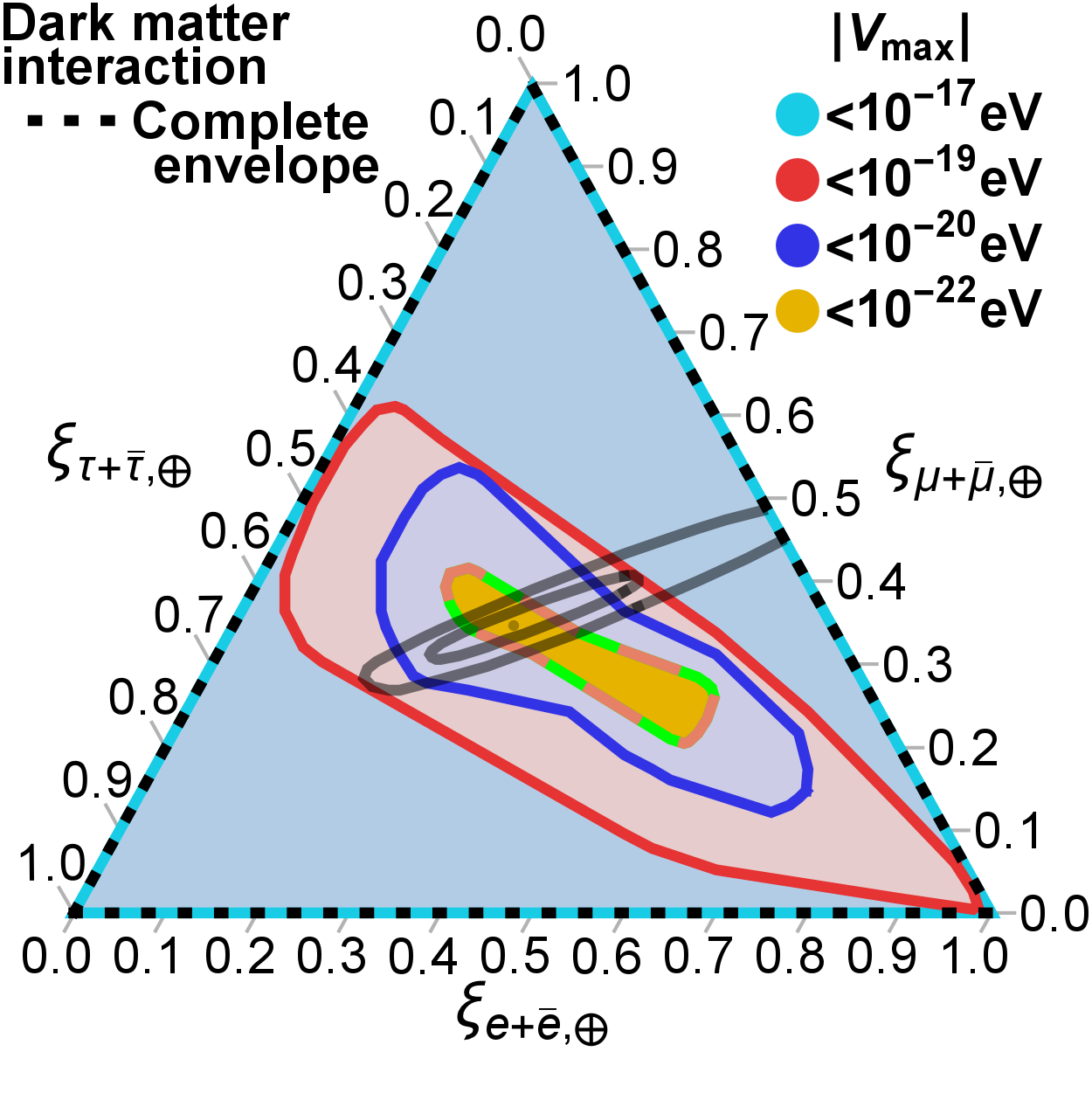} \hspace{0.07\textwidth} 
&
\includegraphics[width=0.4\textwidth]{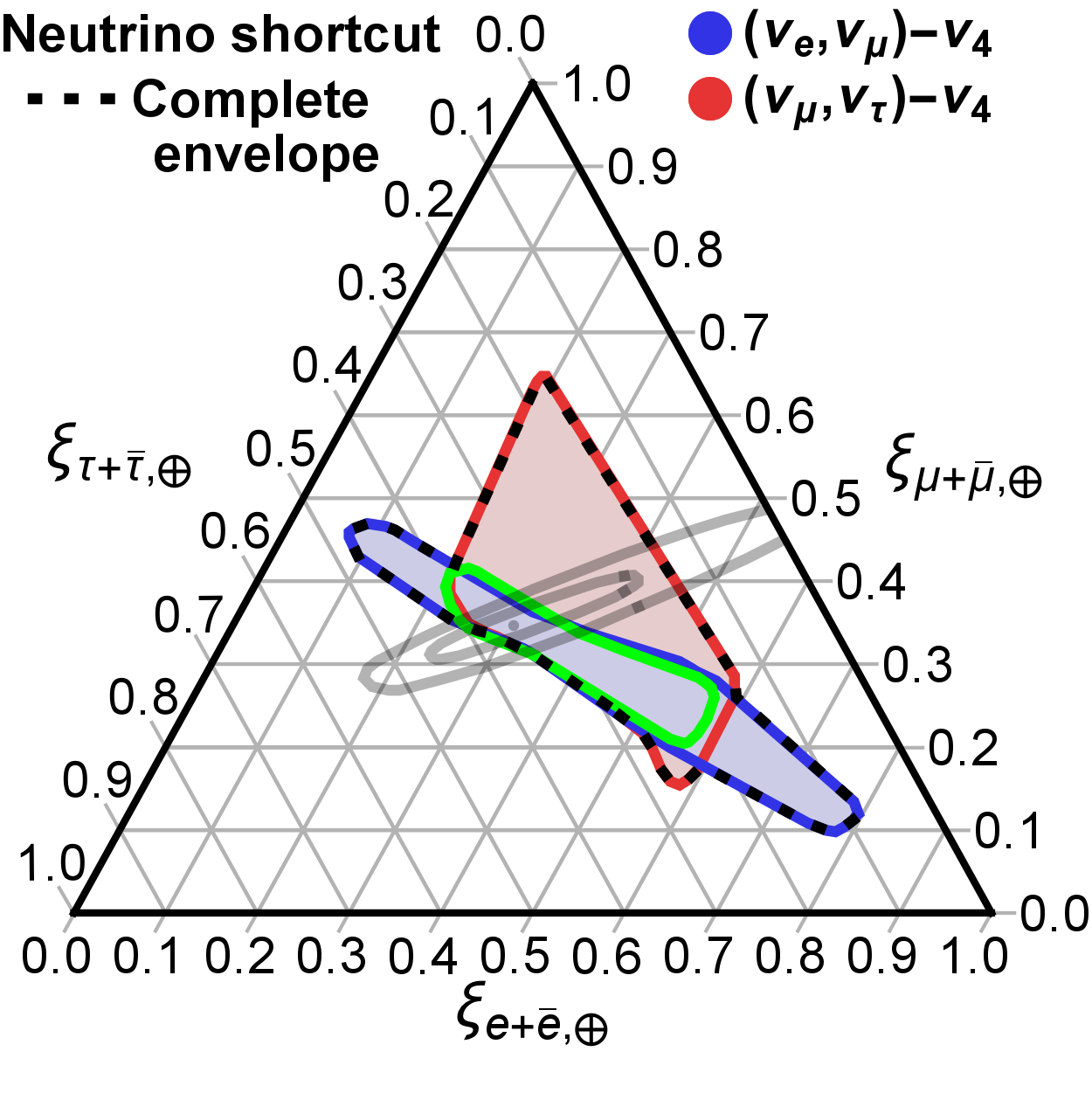}
\end{tabular}
\end{center}
\caption{The allowed flavor compositions for six different BSM scenarios for neutrino propagation. Light green regions mark the standard mixing expectation, gray contours the IceCube-Gen2 expected sensitivity ($1\sigma$, $3\sigma$) for the ``Gen2 scenario''. The best-fit points are everywhere marked by a dot. The dashed contours mark the ``complete envelope'', which is the parameter space in principle allowed -- which is used for reference later.}
\label{fig:propa}
\end{figure*}

In this section, we investigate BSM scenarios affecting the neutrino propagation. Our results are shown in \figu{propa}.

\subsection{Pseudo-Dirac neutrinos}

Due to the astrophysical distance, neutrino telescopes can probe very small mass square differences among neutrinos. This makes them ideal to constrain the pseudo-Dirac neutrino scenario. It has been studied in the general context of neutrino telescopes \cite{Beacom:2003eu, Keranen:2003xd, Esmaili:2009fk}, but also for certain astrophysical sources such as Gamma Ray Bursts (GRBs) \cite{Esmaili:2012ac, Moharana:2013an, Joshipura:2013yba}. Neutrinos might be Dirac or Majorana particles, depending on the allowed mass terms in the SM Lagrangian. The neutrino mass matrix can be written as  
\begin{equation}
 M_{\nu}=
\begin{pmatrix}
 m_{L} & m_{D} \\
m_{D}^{T} & M_{R}
\end{pmatrix},
\end{equation}
where $m_{D}$ is the Dirac mass matrix and $m_{L}, M_{R}$ are the Majorana mass matrices. The mass matrix $m_{L}$ breaks the SM gauge group $SU(2)_{L}$ and should be zero unless new particles (such as a Higgs triplet) are introduced. This allows for a mass term which is invariant under the SM symmetries. Similarly, $M_{R}$ can vanish like $m_{L}$ as a result of new gauge symmetries such as $SU(2)_{R}$ \cite{Langacker:1998ut}, however left-right symmetry models allow $M_{R}$ under this gauge group by introducing a $SU(2)_{R}$ Higgs triplet \cite{Bambhaniya:2013wza, Babu:2014vba, Chakrabortty:2016wkl}. The classical seesaw mechanism \cite{MINKOWSKI1977421, GellMann:1980vs, Yanagida01091980, PhysRevLett.44.912, PhysRevD.22.2227} is generated by assuming $m_{L}=0, m_{D} \ll M_{R}$, whereas $m_{L}=M_{R}=0$ results in pure Dirac neutrinos. The last example leads to six Weyl neutrinos, three pairs of an active and sterile state. Even though the active-sterile mixing angle between the pair is maximal, \ie\ $\theta=\pi/4$, they do not oscillate into their sterile partner since they are degenerate, \ie\ $\delta m^{2}=0$. Lifting this degeneracy, \ie\ $m_{L},M_{R} \ll m_{D}$, leads to the pseudo-Dirac scenario. The mixing angle between the active and sterile states is very close to maximal, $\theta \simeq \pi/4$, and the mass square difference, $\delta m^{2} \simeq 2m_{D}(m_{L}+M_{R})$, is small. The flavor oscillation probability in the pseudo-Dirac scenario is \cite{Beacom:2003eu}
\begin{equation}
 P_{\alpha \beta}=\frac{1}{4}\left|\sum_{j=1}^{3}U_{\alpha j}\left[e^{i(m_{j}^{+})^{2}L/(2E)}+e^{i(m_{j}^{-})^{2}L/(2E)} \right]U_{\beta j}^{*}\right|^{2}
\end{equation}
where $m_{j}^{+}$ ($m_{j}^{-}$) is the heavier (lighter) of the active-sterile neutrino pair $j$, $U$ is the PMNS mixing matrix, $L$ is the baseline and $E$ is the energy. Averaging out $\Delta m_{21}^{2}$ and $\Delta m_{32}^{2}$ (mass square differences among states of different pairs, \ie\ $(\Delta m_{ij}^{+-})^{2} \equiv (m_{i}^{+})^{2}-(m_{j}^{-})^{2}=(\Delta m_{ij}^{--})^{2}+\delta m_{j}^{2} \simeq (\Delta m_{ij}^{--})^{2} \simeq (\Delta m_{ij}^{+-})^{2} \simeq (\Delta m_{ij}^{-+})^{2}$), the oscillation probability becomes 
\begin{align}
 & P_{\alpha \beta}(L, \delta m_{1}^{2}, \delta m_{2}^{2}, \delta m_{3}^{2}) = \nonumber  \\ & \sum_{j=1}^{3} |(U_{\text{PMNS}})_{\alpha j}|^{2}|(U_{\text{PMNS}})_{\beta i}|^{2}\left[ 1-\text{sin}^{2}\left(\frac{\delta m_{j}^{2}L}{4E} \right)\right]
\label{eq:oscpseudo}
\end{align}  
where $\delta m_{j}^{2}=(m_{j}^{+})^{2}-(m_{j}^{-})^{2}$. The mass square differences $\delta m_{j}^{2}$ and the baseline $L$ are unknown, but we average over the baseline and scan for $\delta m_{j}^{2}$ to obtain the flavor mixing 
\begin{equation}
 \bar{P}_{\alpha \beta}=\int_{0}^{L}P_{\alpha \beta}(L^{\prime},\delta m_{1}^{2}, \delta m_{2}^{2},\delta m_{3}^{2})\frac{dL^{\prime}}{L}.                                                                                                                                                                                                                                                                                                                                                               
\end{equation}
This is justified because the sources of IceCube's astrophysical neutrinos are unresolved and $\delta m_{j}^{2}$ can be any value (treated as a continuous theory parameter). We require $L$ to obey $L \gg L_{\text{coh}} \simeq 2E/\Delta m_{21}^{2}$ where $L_{\text{coh}}$ is the coherence distance such that the ordinary mass square differences ($\Delta m_{21}^{2}, \Delta m_{32}^{2}$) are averaged out (requirement for \equ{oscpseudo}). We choose $E \in [10, 10^{4}]$~TeV and $\delta m_{j}^{2} \in [10^{-17}, 10^{-19}]$~eV$^{2}$, the range neutrino telescopes can probe \cite{Esmaili:2009fk}. There are $2^{3}=8$ different possibilities since we can either average over the mass square differences or set it equal to zero. If all are equal to zero, it will lead to standard mixing. Therefore, this is omitted in the end. The seven remaining cases are referred to by the disk notation (except one): a filled disk means that the corresponding $\sin^2=0$, whereas an unfilled disk means that we average it. For example,  ``\makebox{{\makebox[0pt][l]{\ding{182}}} \hspace*{0.08cm} {\makebox[0pt][l]{\ding{173}}} \hspace*{0.08cm} {\makebox[0pt][l]{\ding{174}}}\hspace*{0.27cm}}'' means that 
$\sin^{2} ( \delta m_{1}^{2} L /(4E))=0$, and the other two $\sin^{2} \neq 0$. The disk notation is not shown for scenario ``\makebox{{\makebox[0pt][l]{\ding{172}}} \hspace*{0.08cm} {\makebox[0pt][l]{\ding{173}}} \hspace*{0.08cm} {\makebox[0pt][l]{\ding{174}}}\hspace*{0.27cm}}'', however it generates the complete envelope since all three mass square differences are averaged.   

Our results are shown in the upper left panel of \figu{propa}, including the envelope of all possible scenarios. The figure demonstrates that large deviations from the standard mixing can be expected. 

\subsection{Neutrino decay}

If neutrinos are unstable, they will decay during their propagation to Earth. The equation describing the invisible decay (decay into invisible decay products), which can be incomplete, is \cite{Beacom:2002vi}
\begin{equation}
 P_{\alpha \beta}=\sum_{i=3}^{3}|(U_{\text{PMNS}})_{\alpha i}|^{2}|(U_{\text{PMNS}})_{\beta i}|^{2} \,  \text{exp}\left(-\frac{L m_{i}}{E \tau_{i}}\right),
\label{eq:oscdecay}
\end{equation}
where $\tau_{i}$ is the rest frame lifetime of the $\nu_{i}$ mass eigenstate boosted by $\gamma=E/m_{i}$ into the laboratory frame. For additional studies considering neutrino decay, see \Refs~\cite{Baerwald:2012kc, Bustamante:2016ciw}. The neutrino mass $m_{i}$ is unknown, whereas $L$ and $E$ are experimentally measured. Therefore, one quotes $\tau_{i}/m_{i}$ as the neutrino lifetime. The current limits are: $\tau_{1}/m_{1} \gtrsim 10^{5}$~s/eV for $\nu_{1}$ from SN1987A \cite{PhysRevLett.58.1490}, $\tau_{2}/m_{2} \gtrsim 10^{-4}$~s/eV for $\nu_{2}$ from solar neutrinos \cite{Joshipura:2002fb, Bandyopadhyay:2002qg, Eguchi:2003gg, Aharmim:2004uf}, and $\tau_{3}/m_{3} \gtrsim 10^{-10}$~s/eV for $\nu_{3}$ from atmospheric and long-baseline data \cite{GonzalezGarcia:2008ru}. Neutrino telescopes can probe lifetimes of $\tau/m \gtrsim 10^{2} \frac{L}{\text{Mpc}}\frac{\text{TeV}}{E}$~s/eV \cite{Mehta:2011qb}, which are about a factor of $10^{5}$ longer than the current limits for $\nu_{2}$ and $\nu_{3}$. 

However, \equ{oscdecay} has a shortcoming, namely the possibility of the heavier mass eigenstates repopulating the lighter mass eigenstates. Therefore, we will include this effect in our approach. The total Hamiltonian is given by \cite{Berryman:2014yoa}
\begin{align}
 \mathcal{H}_{\text{tot}}&=\frac{1}{2E}U_{\text{PMNS}}\left[
\begin{pmatrix}
 0 & 0 & 0 \\
 0 & \Delta m_{21}^{2} & 0 \\
0 & 0 & \Delta m_{32}^{2}
\end{pmatrix} \right. \nonumber \\ 
&+ \left. i
\begin{pmatrix}
  -\lambda_{1} & \lambda_{2}\text{Br}_{2 \rightarrow 1} & \lambda_{3}\text{Br}_{3 \rightarrow 1} \\
 0 & -\lambda_{2} & \lambda_{3}\text{Br}_{3 \rightarrow 2} \\
 0 & 0 & -\lambda_{3}
\end{pmatrix}
 \right]U_{\text{PMNS}}^{\dagger}
\label{eq:decayHamiltonian}
\end{align}
where the first term is the usual mass term and the second term is due to decay. Here $\lambda_{i}=m_{i}/\tau_{i}$ and $\text{Br}_{j \rightarrow i}$ is an effective branching ratio between the $j$ and $i$ mass eigenstate. The matrix is upper triangular due to kinematics (lighter mass eigenstates cannot decay into heavier mass eigenstates). We chose $\Delta m_{21}^{2} \in [7.03, 8.09]\cdot 10^{-5}$~eV, $\Delta m_{32}^{2} \in [2.41,2.64]\cdot 10^{-3}$~eV and $E \in [10,10^{4}]$~TeV, and we compute the eigenvalues and eigenvectors by a singular value decomposition (SVD)
\begin{equation}
 \mathcal{H}_{\text{tot}} = V \tilde{\lambda} V^{\dagger}
\end{equation}
where $\tilde{\lambda}=\text{diag}(\tilde{\lambda}_{1}, \tilde{\lambda}_{2}, \tilde{\lambda}_{3})$ with $\tilde{\lambda}_{i}$ being real. The flavor mixing is given by
\begin{equation}
 \bar{P}_{\alpha \beta}=\frac{1}{L}\int_{0}^{L} |\langle \nu_{\beta}|Ve^{-i\tilde{\lambda} L^{\prime}}V^{\dagger}| \nu_{\alpha} \rangle|^{2}dL^{\prime},
\end{equation}
where we average over the distance to include decoherence effects. Therefore, $L$ has to obey $L \gg L_{\text{coh}} \simeq 2E/\Delta m_{21}^{2}$ where $L_{\text{coh}}$ is the coherence distance, however $L$ cannot exceed the Hubble distance, see \Ref~\cite{Baerwald:2012kc} for explanation. There are $2^{3}=8$ possibilities of neutrino decays, since either active state may be stable or not; intermediate unstable states can be integrated out~\cite{Maltoni:2008jr}. We will investigate seven cases since one is trivial, namely the case with all mass eigenstates stable (standard mixing). We again use the disk notation, for instance, ``\makebox{{\makebox[0pt][l]{\ding{182}}} \hspace*{0.08cm} {\makebox[0pt][l]{\ding{173}}} \hspace*{0.08cm} {\makebox[0pt][l]{\ding{174}}}\hspace*{0.27cm}}'' means $\nu_1$ stable and $\nu_2$, $\nu_3$ unstable. Possible scenarios and branchings for the different mass orderings are discussed in \Ref~\cite{Maltoni:2008jr}.

The expected flavor composition is shown in the upper left panel of \figu{propa}, including the complete envelope -- which is identical to the scenario where all mass eigenstates are unstable (for complete decays, one would not expect a flux at Earth). IceCube-Gen2 can exclude 85\% of the envelope area covered by all decay scenarios.  

It is interesting to compare neutrino decay and pseudo-Dirac in this context.  Essentially, \equ{oscpseudo} looks similar to \equ{oscdecay} with $[1-\text{sin}^{2} (\delta m_{i}^{2}L/(4E) ) ]=\text{exp}(-L m_{i})/(E \tau_{i})$. Therefore, scanning over different parameters for the two cases, the same parameter space is obtained. However, we use \equ{decayHamiltonian} for decay, where we marginalize decay rates and branchings. This fact together with the off-diagonal elements in the decay term being smaller than the diagonal elements, \ie\ $|\lambda_{i}|~\geq~|\lambda_{i}\text{Br}_{i \rightarrow j}|$, and averaging the flavor mixing over astrophysical distances, means re-occupations of mass eigenstates are equivalent to use \equ{oscdecay} with lower decay rates. Therefore, one reproduces the same parameter space for the decay and pseudo-Dirac cases, even though we use \equ{decayHamiltonian} instead of \equ{oscdecay}.

\subsection{Quantum decoherence}

Quantum decoherence~\cite{Hooper:2004xr, Morgan:2004vv} can alter the flavor mixing, and it depends only on two non-zero quantum decoherence parameters $\Psi$ and $\Gamma$ \cite{Hooper:2005jp}
\begin{align}
 P_{\alpha \beta}=\frac{1}{3}& + \frac{1}{2}(|U_{\alpha 1}|^{2}-|U_{\alpha 2}|^{2})(|U_{\beta 1}|^{2}-|U_{\beta 2}|^{2})\text{e}^{-2\Psi L E^{n}} \nonumber \\ & +\frac{1}{6}(|U_{\alpha 1}|^{2} +|U_{\alpha 2}|^{2}-2|U_{\alpha 3}|^{2}) \nonumber \\
& \quad \times (|U_{\beta 1}|^{2}+|U_{\beta 2}|^{2}-2|U_{\beta 3}|^{2})\text{e}^{-2\Gamma L E^{n}}
\label{eq:flavormixQD}
\end{align}
under the assumption that oscillations average out over astrophysical distances. The number $n$ carries the energy dependent imprint of a specific model, and the usual values used in the literature are $n=-1,0,2$ \cite{Blennow:2005yk}.  Interestingly, quantum decoherence can therefore occur as a high energy effect such as a remnant of a quantum theory of gravity. We choose the exponential factors between zero and one rather than choosing specific values of $\Psi$, $\Gamma$, $L$ and $E$ to cover the possible model parameter space. We note that this scenario is, in terms of the envelope covered, not different from standard mixing; see \figu{propa} upper right corner.
Quantum decoherence may, however, produce some interesting energy-dependent effects, as we will see later.  

\subsection{Sterile neutrinos}

Many theories beyond the SM include sterile neutrinos. This will affect the flavor composition since the active neutrinos can mix with the sterile neutrinos during their propagation. Using the same method as in \Sec~\ref{sec:NSnuProd} and restricting ourselves to active neutrinos at the source, we find the region in the middle left panel of \figu{propa}. This region is significantly larger than the standard mixing region, and 86\% of it can be excluded with IceCube-Gen2. Note again that here the current bounds on the unitarity of the mixing matrix elements have been used, which means that the region will slightly shrink in the future.

\subsection{Effective operators}

Higher-dimensional terms may originate from a high energy scale theory and lead to new physics via effective operators. We use \Ref~\cite{Arguelles:2015dca} as a guideline, investigating the same examples, however the neutrino oscillation parameters obey the constraints derived from the $\chi^{2}$ (\equ{chisquare}). The Hamiltonian becomes 
\begin{align}
 H_{\text{tot}}&=\frac{1}{2E}U_{\text{PMNS}}\text{diag}(0,\Delta m_{21}^{2}, \Delta m_{32}^{2})U_{\text{PMNS}}^{\dagger}  \nonumber \\
& +\sum_{n}\left(\frac{E}{\Lambda_{n}}\right)^{n}\tilde{U}_{n}\mathcal{O}\text{diag}(O_{n,1},O_{n,2},O_{n,3})\tilde{U}_{n}^{\dagger} \nonumber \\
& =V\text{diag}(\Delta_{1},\Delta_{2},\Delta_{3}) V^{\dagger} \, ,
\label{eq:hamiltonianeff}
\end{align}
where $V$ is the mixing matrix that results from diagonalizing $H_{\text{tot}}$ and $O_{n,i} \sim \text{O}(1)$ for $i\in[1,3]$. The new physics depends on the coupling strengths $\mathcal{O}$ and $\Lambda_{n}$, and we choose $\tilde{U}_{n}$ in a parameterization-independent way, meaning we obtain the whole allowed parameter space. We investigate the $n=1$ effective operator's impact on the flavor composition since the lower  terms are more relevant. This is the usual perception when one considers a renormalizable quantum field theory, \ie\ $(E/\Lambda_{n}) \ll 1$ for every $n$. Neutrino oscillation data constrains the coupling strength on the effective operator to $\mathcal{O} \sim 10^{-23}$~GeV \cite{Abe:2014wla, Abbasi:2010kx}. As in \Ref~\cite{Arguelles:2015dca}, we study $\mathcal{O} \sim 10^{-23}$~GeV with $\Lambda_{1}=1$~TeV as current limit of the $n=1$ effective operator. We also study the cases $\mathcal{O} \sim 10^{-26}$~GeV with $\Lambda_{1}=35$~TeV and $\mathcal{O} \sim 10^{-28}$~GeV with $\Lambda_{1}=2$~PeV. The choice of these values makes the new physics to be the same order of magnitude as the mass term in the Hamiltonian with a neutrino energy of $E=35$~TeV and $E=2$~PeV, respectively \cite{Arguelles:2015dca}. The flavor mixing is $P_{\alpha \beta}=\sum_{i=1}^{3}|V_{\alpha i}|^{2}|V_{\beta i}|^{2}$, now depending on $V$ rather than $U_{\text{PMNS}}$ since $V$ diagonalizes the Hamiltonian (\equ{hamiltonianeff}).

The parameter space for different values of $O$ is shown in the middle right panel of \figu{propa}. Since we allow for arbitrary initial flavors excluding tau neutrinos, it can cover almost the whole plane except from the $\nu_\tau$ corner. If one allows for tau neutrinos at the source, the lower left corner is probed \cite{Arguelles:2015dca}. IceCube-Gen2 can exclude 94\% of the parameter space for $O \simeq 10^{-23} \, \mathrm{GeV}$.

\subsection{Interaction with dark matter}

We follow the approach in \Ref~\cite{deSalas:2016svi} for this section. Neutrinos are produced by the cosmic accelerators, and IceCube has discovered an isotropic neutrino flux \cite{Aartsen:2014muf}. As neutrinos propagate from the source towards Earth, they might interact with dark matter (DM). There are different models describing the DM distribution in the Universe. The effect of neutrino-DM interaction introduces a potential in the Hamiltonian
\begin{equation}
 \mathcal{H}_{\text{tot}}=\frac{1}{2E}U_{\text{PMNS}}
\begin{pmatrix}
 0 & 0 & 0 \\
0 & \Delta m_{21}^{2} & 0 \\
0 & 0 & \Delta m_{32}^{2}
\end{pmatrix}
U_{\text{PMNS}}^{\dagger}+\mathcal{V}
\end{equation}
with the mass square differences $\Delta m_{ij}^{2}$ and the neutrino mixing matrix $U_{\text{PMNS}}$. The potential $\mathcal{V}$ describes the interaction between neutrinos and DM, and can be parameterized as \cite{deSalas:2016svi}
\begin{equation}
 \mathcal{V}_{\alpha \beta}=\lambda_{\alpha \beta}G_{F}N_{\chi}
\end{equation}
where $\lambda_{\alpha \beta}$ is a Hermitian matrix containing the $\pm$~O(1) coupling between neutrinos and DM, the Fermi constant $G_{F}$, and the dark matter number density $N_{\chi}$. The number density is related to the energy density by 
\begin{equation}
 N_{\chi}=\frac{\rho_{\text{DM}}}{m_{\text{DM}}},
\end{equation} and we will use the Navarro-Frenk-White (NFW) DM distribution for $\rho_{\text{DM}}$. It is given by $\rho_{\text{NFW}}(r)=\rho_{\text{DM}}(r, \text{20 kpc}, 1, 3, 1)$ \cite{Navarro:1995iw, Navarro:1996gj} with the general parameterization
\begin{equation}
 \rho_{\text{DM}}(r, r_{s}, \alpha, \beta, \gamma)=\rho_{\oplus} \left(\frac{r_{\oplus}}{r}\right)^{\gamma}\left( \frac{1+(r_{\oplus}/r_{s})^{\alpha}}{1+(r/r_{s})^{\alpha}}\right)^{(\beta-\gamma)/\alpha}.
\end{equation}
It is derived under a spherical symmetry and the Galactic Center (GC) corresponds to $r=0$. We assume the solar system being $r_{\oplus}=8.5$~kpc from the GC and several studies shows that $\rho_{\oplus}=0.4 \ \text{GeV}/\text{cm}^{3}$ \cite{Catena:2009mf, Pato:2015dua, doi:10.1093/mnras/stw565}. The distance from the GC to the neutrino position is given by $r^{2}=r_{\oplus}^{2}+l^{2}-2r_{\oplus}l\cos\phi$ where $l=\sum_{i=1}^{n}l_{i}$ is the line of sight distance from the solar system. The specific timestep is given by the index $i$, whereas $n$ defines the total number of timesteps. The ranges are $\phi \in [\pi/2, \pi]$ and $l \in [0,20]$~kpc, ensuring that the effects of mass square differences will average out. The flavor mixing will depend on these two quantities
\begin{align}
 P_{\alpha \beta} &=|\langle \nu_{\beta}| \Pi_{i=1}^{n}U_{f}(l_{i},\phi)|\nu_{\alpha} \rangle|^{2} \nonumber \\
&=|\langle \nu_{\beta}| \Pi_{i=1}^{n}U(l_{i},\phi)e^{-i\mathcal{H_{\text{diag}}}l_{i}}U^{\dagger}(l_{i},\phi)|\nu_{\alpha} \rangle|^{2}.
\end{align}
where $U_{f}(l_{i}, \phi)=U(l_{i},\phi)e^{-i\mathcal{H_{\text{diag}}}l_{i}}U^{\dagger}(l_{i},\phi)$ is the transportation matrix and $\mathcal{H_{\text{diag}}}=U^{\dagger}(l_{i},\phi)\mathcal{H}_{\text{tot}}U(l_{i},\phi)=\frac{1}{2E}\text{diag}(0,(\Delta m_{21, \text{eff}}^{2})_{i}, (\Delta m_{32, \text{eff}}^{2})_{i})$. Therefore, one has to diagonalize the Hamiltonian for every timestep. However, if the propagation is assumed to be adiabatic, the flavor mixing simplifies, and it becomes
\begin{equation}
 P_{\alpha \beta}=|\langle \nu_{\beta}|U(0,\phi)e^{-i\mathcal{H_{\text{diag}}}l}U^{\dagger}(l,\phi)|\nu_{\alpha} \rangle|^{2}.
\end{equation}
The adiabatic assumption has been checked in \Ref~\cite{deSalas:2016svi}. By scanning for different distances and angles, the allowed flavor compositions can be computed, and the result is shown in \figu{propa} (lower left panel) for various coupling strengths between the neutrinos and dark matter. This effect is unique among those scenarios we studied, since it covers the entire flavor triangle.

Phenomenologically speaking, this scenario looks similar to the effective operator case. One adds a potential to the Hamiltonian, but they are realizing different percentages of the flavor triangle: $\sim 100$\% for the $\nu$-DM interaction scenario vs. $\sim 68$\% for the effective operator case. This is a consequence of the potential in the DM scenario which varies as a function of the DM density, whereas the potential is constant for the effective operator case. If the potential in the DM scenario was constant, the parameter space would be the same as for the effective operator parameter space. 

This scenario depends on the arrival direction which, in turn, describes the DM density in the Universe. Looking toward the GC rather than away, one expects larger deviations in the flavor composition due to the larger density of DM. Therefore, one can (in principle) discriminate this model by investigating the arrival direction. 

\subsection{Sterile neutrino shortcut through extra dimension}

Theories beyond the SM with extra dimensions confines the SM particles on a four dimensional (time + spatial dimensions) brane embedded in an extra-dimensional bulk \cite{Pas:2005rb, Esmaili:2014esa, 0954-3899-40-5-055202, 0954-3899-40-7-079501, Kisselev2010, Lykken:2007kp}. Any singlet particle, \ie\ sterile neutrino, may travel off the brane since they are not confined on the brane by the SM's symmetries. Therefore, the line element is given by \cite{Chung:1999xg,Chung:1999zs,Csaki:2000dm}
\begin{equation}
 ds^{2}=dt^{2}-\sum_{i=1}^{3}\eta^{2}(u)(dx^{i})^{2}-du^{2}.
\end{equation}
where $u$ is the extra dimension and $\eta^{2}(u)$ describes the brane's embedding in the bulk. If the brane is flat in its embedding, then the sterile neutrino geodesic is the same as the active neutrino geodesic. However, if the brane is curved in its embedding, then the sterile neutrino may have a different trajectory. As a consequence, the dispersion relation is altered and the sterile neutrino will experience a shorter propagation time. Therefore, the effective two-flavor mixing angle becomes \cite{Pas:2005rb}
\begin{equation}
 \text{sin}^{2}(2\tilde{\theta})=\frac{\text{sin}^{2}(2\theta)}{\text{sin}^{2}(2\theta)+\text{cos}^{2}(2\theta)\left[1-\left(\frac{E}{E_{\text{res}}} \right) \right]^{2}}
\label{eq:twoflavormixangle}
\end{equation}
where $E_{\text{res}}$ denotes the resonance energy
\begin{equation}
 E_{\text{res}}=\sqrt{\frac{\delta m^{2}\cos(2\theta)}{2\epsilon}}.
\end{equation}
Here $\delta m^{2}$ is the mass square difference in vacuum between the sterile and active neutrino state, $\theta$ is the active-sterile mixing angle in vacuum and $\epsilon$ is the shortcut parameter, $\epsilon=\delta t/t$, defined as the normalized difference of propagation times on the brane and in the bulk. For energies much smaller than the resonance energy $E_{\text{res}}$, the change in the dispersion relation is small, whereas for energies much larger, the effective active-sterile mixing angle is suppressed. Therefore, $E \gg E_{\text{res}}$ and $E \ll E_{\text{res}}$ resembles standard mixing if no significant mixing between the active and sterile state occurs. This also means the PMNS mixing matrix will be unitary when $E \gg E_{\text{res}}$ and $E \ll E_{\text{res}}$, whereas it becomes non-unitary for $E \rightarrow E_{\text{res}}$. We will, with this in mind, compute the flavor composition at $E=E_{\text{res}}$ to investigate the maximal impact for this scenario. 

We follow \Ref~\cite{Aeikens:2014yga} here, which assumes that the active-sterile mixing angles in a four neutrino framework are described by \equ{twoflavormixangle}, and we adopt their scenarios only. We pick $E=E_{\text{res}}$ to study the maximal impact on the flavor composition, meaning the active-sterile mixing angles are $\tilde{\theta}=45^{\circ}$. If $E\neq E_{\text{res}}$, then $\tilde{\theta}$ is computed by \equ{twoflavormixangle} and depends on $\theta$, similar to \Ref~\cite{Aeikens:2014yga}. However, we use \equ{chisquare} to constrain the neutrino oscillation parameters rather than assuming $\sin^{2}\theta_{12}=1/3$, $\sin^{2}\theta_{23}=\pi/4$ and $\sin^{2}\theta_{13}=0$, and we include CP-violation. The mixing matrix is parameterized as $U_{4\times4}=U_{23}\bar{U}_{13}U_{12}\bar{U}_{14}\bar{U}_{24}U_{34}$ where $U_{ij} \ (\bar{U}_{ij})$ is a real (complex) rotation matrix in the $ij$-plane. We investigate two cases taken from \Ref~\cite{Aeikens:2014yga}, where $\tilde{\theta}$ refers to the mixing angle including the extra dimension shortcut: 
\begin{itemize}
 \item Sterile neutrino mixes with the electron and muon neutrino at the same strength, \ie\ $\tilde{\theta}_{14}=\pi/4$, $\tilde{\theta}_{24}=\pi/4$ and $\tilde{\theta}_{34}=0$
\item Sterile neutrino mixes with the muon and tau neutrino at the same strength, \ie\ $\tilde{\theta}_{14}=0$, $\tilde{\theta}_{24}=\pi/4$ and $\tilde{\theta}_{34}=\pi/4$
\end{itemize}
Our results are shown in the lower right panel of \figu{propa}.  The cases $(\nu_{e}, \nu_{\mu})-\nu_{4}$ and $(\nu_{\mu}, \nu_{\tau})-\nu_{4}$ exhibit large deviations from standard mixing. IceCube-Gen2 can exclude 80\% of the combined parameter space of these two cases. We also tested effective maximal mixing between sterile neutrino and all of the active ones, which produces even larger deviations of the flavor composition; however we could not make it evident that this effective scenario can be implemented in a four flavor framework, given all constraints. 

\section{Detection effects}
\label{sec:detectioneffects}
We investigate effects at detection or close to the detector in this section. 

\begin{figure*}[t]
\centering
\begin{tabular}{c c}
\includegraphics[width=0.45\textwidth]{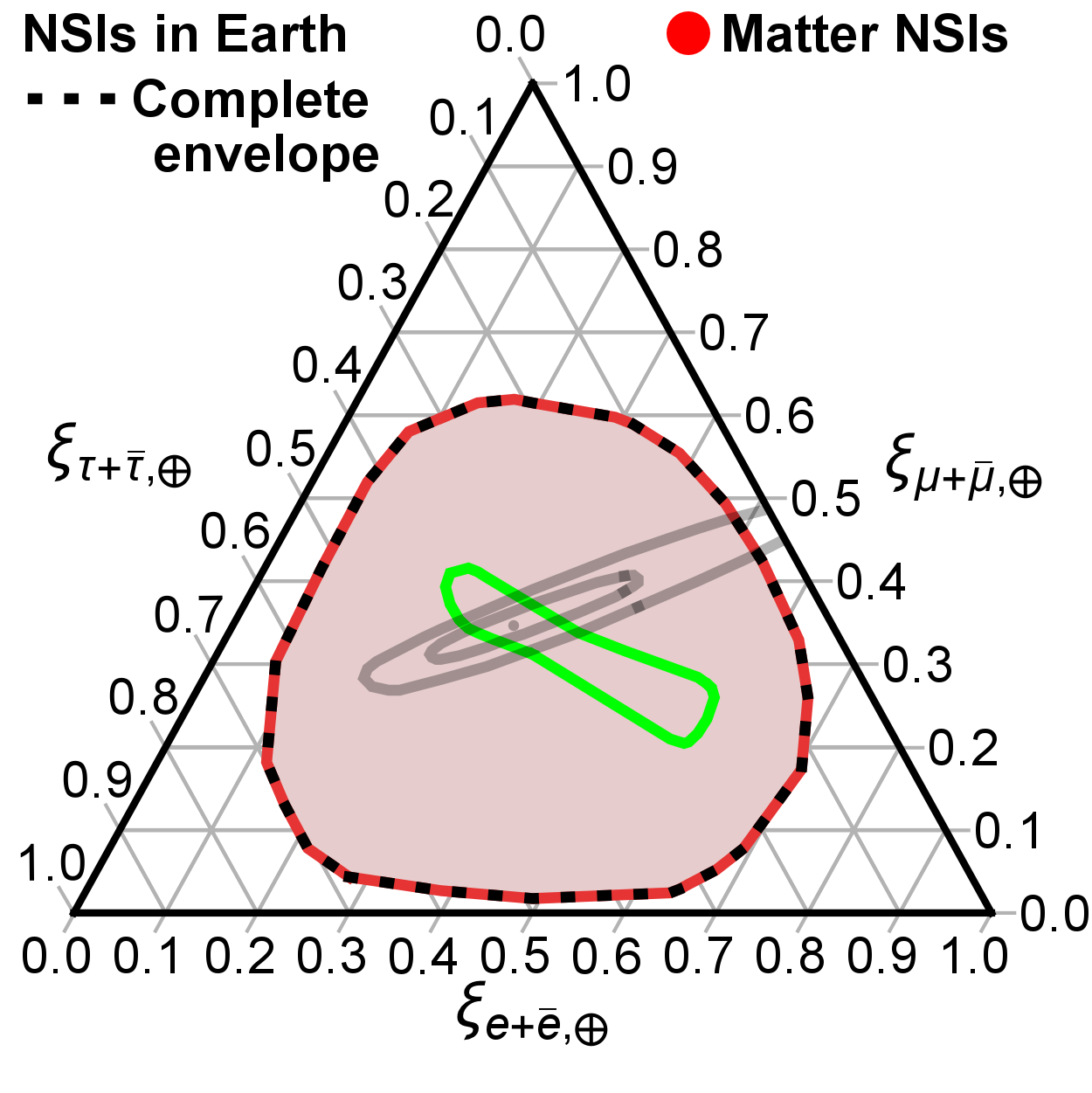}\hspace{0.07\textwidth} &
\includegraphics[width=0.45\textwidth]{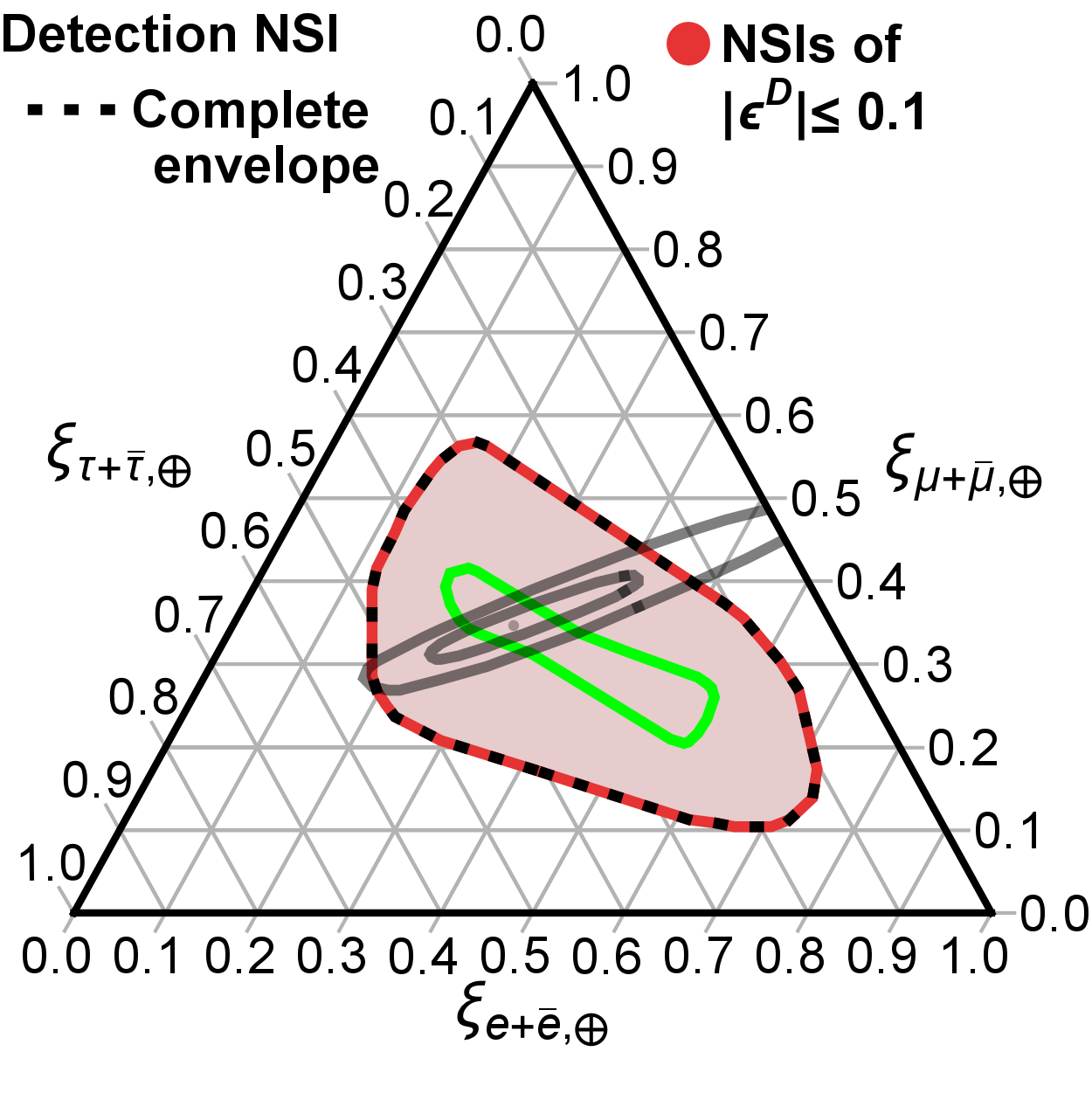}
\\
\end{tabular}
\caption{The allowed flavor compositions for non-standard interactions in Earth matter (left panel) or at detection (right panel). Green regions mark the standard mixing expectation, gray contours the IceCube-Gen2 expected sensitivity ($1\sigma$, $3\sigma$) for the ``Gen2 scenario''. The best-fit points are everywhere marked by a dot. The dashed contours mark the ``complete envelope'', which is the parameter space in principle allowed -- which is used for reference later.}
\label{fig:detection}
\end{figure*}

\subsection{Non-standard interactions in Earth matter}

If neutrinos travel through Earth matter before they are detected, non-standard neutrino interactions can change the flavor composition. We use \Ref~\cite{Gonzalez-Garcia:2016gpq} as a guideline for this scenario. The most general matter Hamiltonian with NSIs is given by \cite{Gonzalez-Garcia:2013usa}
\begin{equation}
 H_{\text{mat}}^{\text{NSI}}=\sqrt{2}G_{F}N_{e}(r)
\begin{pmatrix}
 1+\epsilon_{ee} & \epsilon_{e\mu} & \epsilon_{e\tau} \\
 \epsilon_{e\mu}^{*} & \epsilon_{\mu \mu} & \epsilon_{\mu \tau} \\
\epsilon_{e\tau}^{*} & \epsilon_{\mu \tau}^{*} & \epsilon_{\tau \tau}
\end{pmatrix},
\end{equation}
where $G_{F}$ is the Fermi coupling, $N_{e}(r)$ is the electron density at distance $r$ and $\epsilon_{\alpha \beta}$ are dimensionless parameters encoding the deviation from standard interactions. They are given by \cite{Gonzalez-Garcia:2013usa}
\begin{equation}
 \epsilon_{\alpha \beta}=\epsilon_{\alpha \beta}^{e} + Y_{u} \epsilon_{\alpha \beta}^{u}  + Y_{d} \epsilon_{\alpha \beta}^{d} 
\end{equation}
where $Y_{u}=3.051$ $(Y_{d}=3.102)$ is the average up-quark/electron (down-quark/electron) ratio in the Preliminary Reference Earth Model (PREM) \cite{DZIEWONSKI1981297}. Here $\epsilon_{\alpha \beta}^{e}, \epsilon_{\alpha \beta}^{u}, \epsilon_{\alpha \beta}^{d}$ are the individual NSIs involving the electron, up-quark and down-quark, respectively. We  omit the electron NSI $(\epsilon_{\alpha \beta}^{e})$ from our analysis since it enters both in the complete NSI $\epsilon_{\alpha \beta}$ and in the neutrino cross section. Therefore, it can be difficult to distinguish the new physics from the cross section and the matter potential. In addition, we assume the individual NSIs to be real. The current constraints on $\epsilon_{\alpha \beta}^{u}$ and $\epsilon_{\alpha \beta}^{d}$ are summarized in 
\Ref~\cite{Gonzalez-Garcia:2013usa}, which we vary in their $3 \sigma$ allowed ranges. The flavor mixing over astrophysical distances from the source to the detector is given by \Ref~\cite{Gonzalez-Garcia:2016gpq}, which we marginalize over all possible trajectories through Earth matter.

Our result is shown in \figu{detection}, left panel. In this case, it seems that a relatively large region of the parameter space can be covered, which however depends on the trajectory through Earth matter. Therefore the IceCube-Gen2 allowed region has to be interpreted in a zenith-angle-dependent way, which, while beyond the scope of this work, would in principle allow to distinguish this scenario from others. However, the example illustrates that non-standard interactions in Earth matter can alter the flavor composition substantially, which can be checked by comparing the neutrino flux from different directions (such as up-going versus down-going).

\subsection{Non-standard interactions at detection}

We consider NSIs at detection, meaning the neutrino states are \cite{Ohlsson:2012kf}
\begin{align}
 |\nu_{\alpha}^{P} \rangle & = U_{\text{PMNS}}|\nu_{i} \rangle ,\\
\langle \nu_{\beta}^{D}| &= \langle \nu_{i}|U_{\text{PMNS}}^{\dagger}(1+\epsilon^{D})^{\dagger}
\end{align}
where $|\nu_{\alpha}^{P} \rangle$ ($\langle \nu_{\beta}^{D}|$) represents the neutrino at production (detection), $|\nu_{i} \rangle$ is the mass eigenstate, $U_{\text{PMNS}}$ is the PMNS mixing matrix and $\epsilon^{D}$ represents the NSIs at detection. The flavor mixing is $P_{\alpha \beta}=\sum_{i}|\mathcal{J}_{\alpha \beta}^{i}|^{2}$ with
\begin{align}
 \mathcal{J}_{\alpha \beta}^{i}= &(U_{\text{PMNS}})_{\alpha i}^{*}(U_{\text{PMNS}})_{\beta i} \nonumber \\
& +\sum_{\gamma}\epsilon^{D}_{\gamma \beta}(U_{\text{PMNS}})_{\alpha i}^{*}(U_{\text{PMNS}})_{\gamma i}
\end{align}
(note the difference in indices compared to NSIs at production). We use the same benchmark value for the detection NSI as for the production NSI, \ie\ $|\epsilon^{D}| \leq 0.1$, and the neutrino oscillation parameters are constrained by $\chi^{2}$. The impact of detection NSIs on the flavor composition is shown in \figu{detection}, right panel. Evidently from the figure, the detection NSI parameter space is larger than that of production NSI. The production effect is similar to a different flavor composition at the source, which averages out over astrophysical distances. Therefore, production NSI impact the flavor composition less than detection NSI. IceCube-Gen2 can exclude 89\% of the allowed parameter space.

\section{Discrimination by flavor}
\label{sec:flavordiscrimination}

\begin{figure*}[tp]
\centering
\begin{tabular}{c c}
\includegraphics[width=0.32\textwidth]{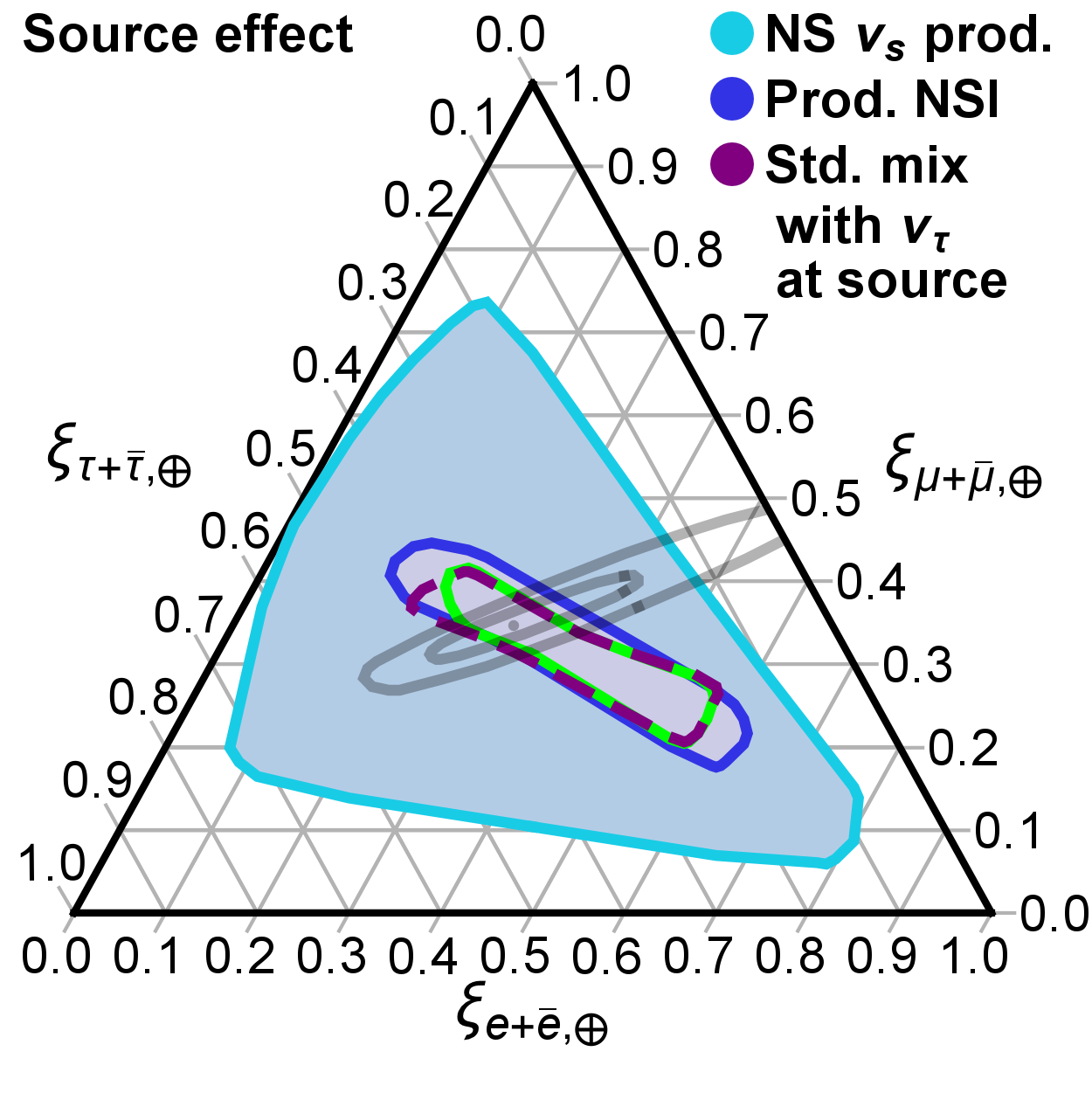}\hspace{0.01\textwidth}
\includegraphics[width=0.32\textwidth]{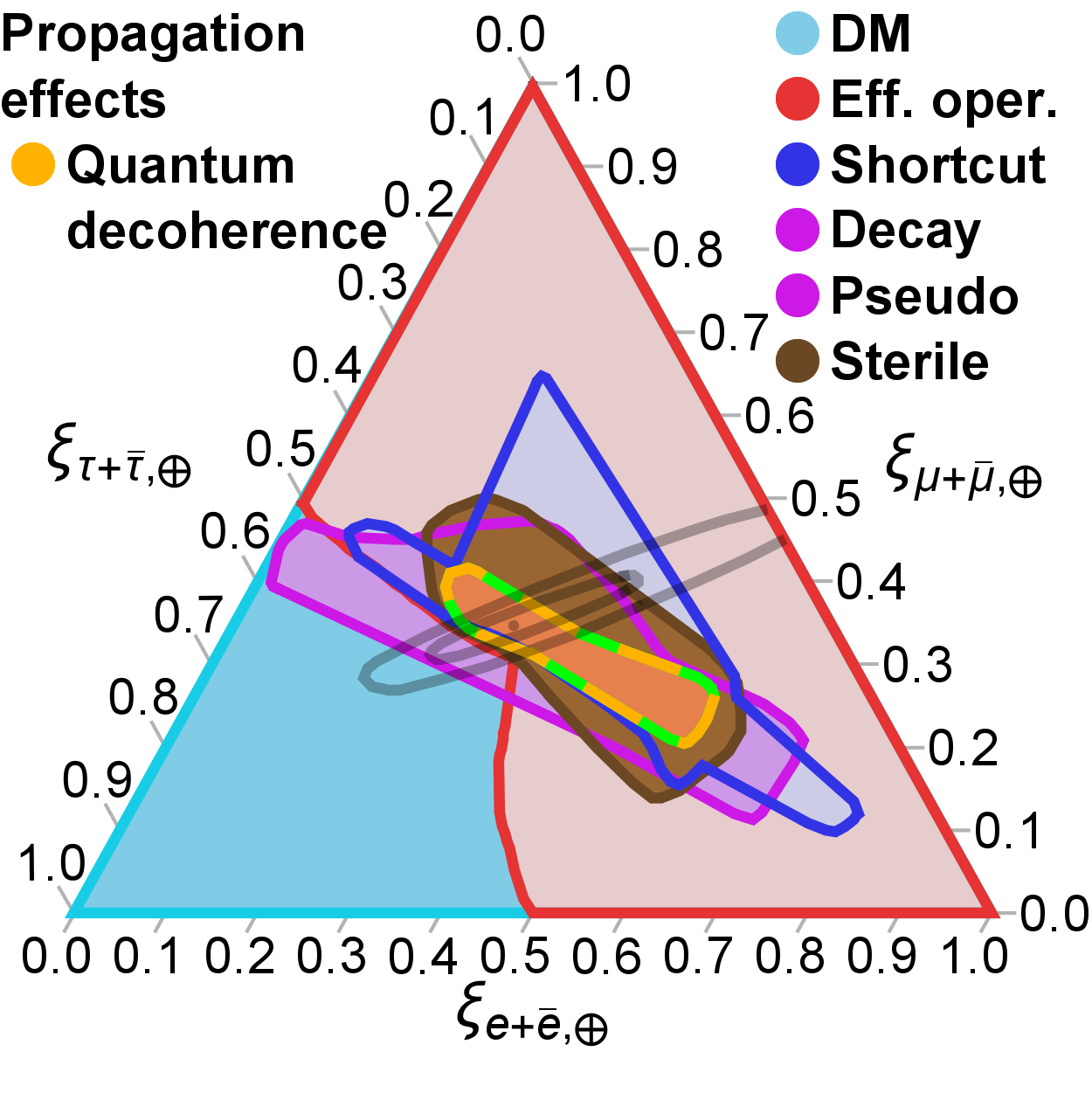}\hspace{0.01\textwidth}
\includegraphics[width=0.32\textwidth]{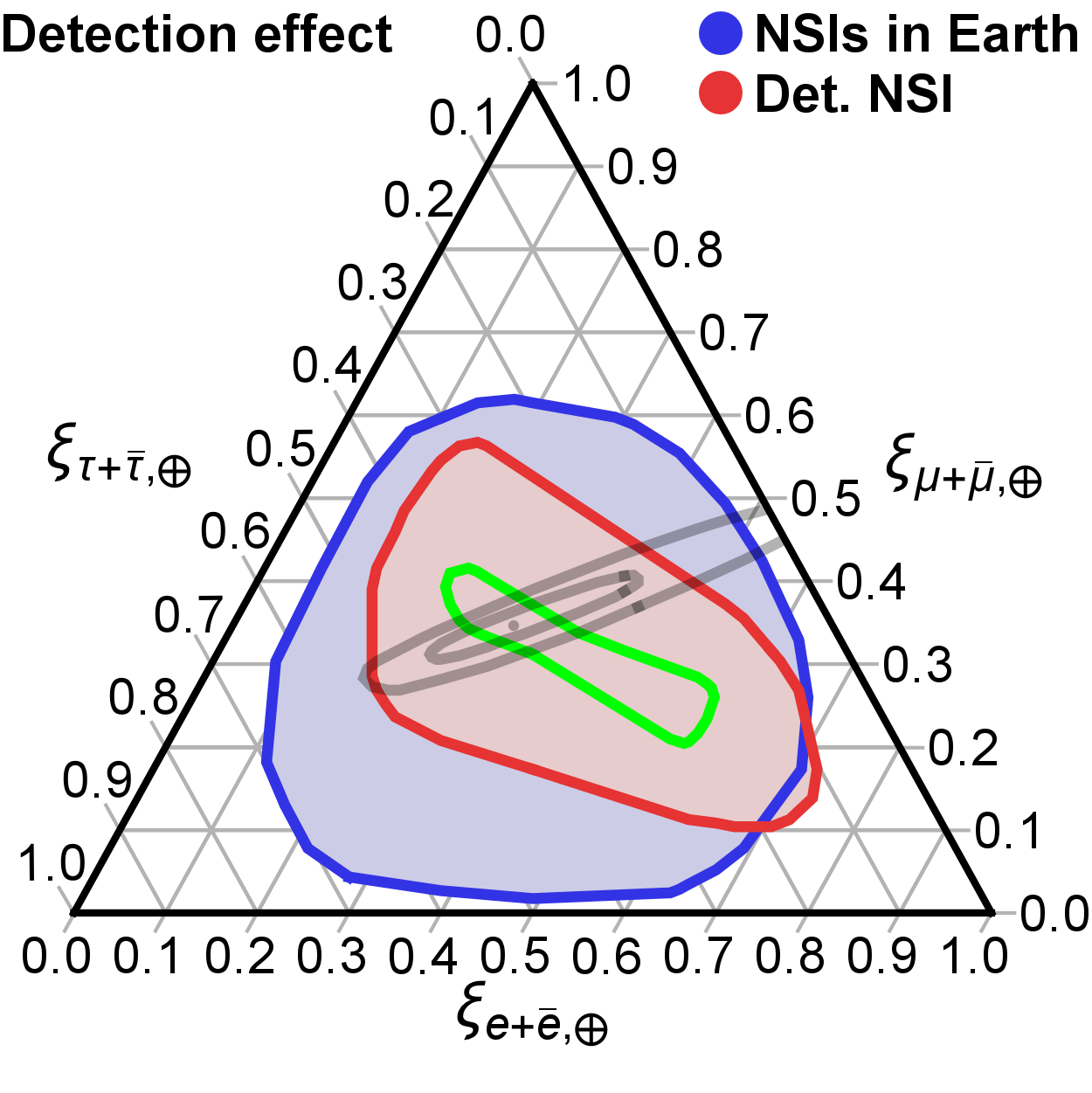}
\end{tabular}
\caption{Allowed parameter space for selected source (left panel), propagation (middle panel), and detection BSM effects (right panel), where we show the complete envelopes for the allowed parameter spaces. Green regions mark the standard mixing expectation, gray contours the IceCube-Gen2 expected sensitivity ($1\sigma$, $3\sigma$) for the ``Gen2 scenario''. The best-fit points are everywhere marked by a dot.
}
\label{fig:comparison}
\end{figure*}

Here we discuss how well the flavor composition can discriminate among different BSM scenarios.

We show a comparison of the allowed parameter space for certain source (left panel), propagation (middle panel), and detection (right panel) BSM effects in \figu{comparison}. From the figure, one can read off that $\nu_s$ produced at the source, dark matter interactions or effective operators relevant during propagation, or non-standard interactions in Earth matter produce potentially large deviations from standard mixing. Interestingly, the lower left corner of the triangle can only be reached by dark matter interactions. Note that some effects, such as non-standard interactions in Earth matter and dark matter interactions, can be potentially identified by comparing different arrival directions of the neutrinos. 

\begin{table}[t]
\centering
  \setlength{\tabcolsep}{2pt}
  \begin{tabular}{p{3.5cm}|p{2cm}|p{2cm}}
      \hline
      Scenario  & Exclusion by IceCube & Exclusion by IceCube-Gen2  \\
      \hline 
      Complete flavor triangle & \hspace*{0.5cm} 42\% & \hspace*{0.5cm} 96\%
      \\
      Standard mixing & \hspace*{0.55cm} 2\% & \hspace*{0.5cm} 73\%
      \\
      Non-standard neutrino production  & \hspace*{0.5cm} 17\%  & \hspace*{0.5cm} 93\%
      \\
      NSI at production & \hspace*{0.55cm} 5\% & \hspace*{0.5cm} 84\%
      \\
      Matter effetcs & \hspace*{0.55cm} 0\% & \hspace*{0.5cm} 71\%
      \\
      Pseudo-Dirac neutrino & \hspace*{0.5cm} 14\% & \hspace*{0.5cm} 85\%
      \\
      Decay & \hspace*{0.5cm} 14\% & \hspace*{0.5cm} 85\%
      \\
      Quantum decoherence & \hspace*{0.55cm} 2\% & \hspace*{0.5cm} 73\%
      \\
      Sterile neutrino & \hspace*{0.5cm} 10\% & \hspace*{0.5cm} 86\%
      \\
      Effective operator & \hspace*{0.5cm} 36\% & \hspace*{0.5cm} 94\%
      \\
      Interaction with DM & \hspace*{0.5cm} 42\% & \hspace*{0.5cm} 96\%
      \\
      Shortcut through extra dimension & \hspace*{0.5cm} 11\% & \hspace*{0.5cm} 80\%
      \\
      NSI in Earth matter & \hspace*{0.5cm} 30\% & \hspace*{0.5cm} 92\%
      \\
      NSI at detection & \hspace*{0.5cm} 11\% & \hspace*{0.5cm} 89\%
      \\
      \hline
  \end{tabular}
  \caption{Percent of parameter space for the BSM scenarios and matter effects excludable at $3\sigma$ by IceCube and IceCube-Gen2. We included IceCube for comparison to IceCube-Gen2. We have taken the complete envelope as the parameter space, and we have not considered the individual sub-parameter spaces.}
 \label{tab:Gen2exlcusion}
\end{table}

We quantify the parameter space exclusion by IceCube and IceCube-Gen2 in \tabl{Gen2exlcusion}. Even at $3\sigma$ CL, IceCube can exclude $42$\% of the parameter space, whereas it is $96$\% for IceCube-Gen2. A few examples of BSM physics with a high exclusion percentage (more than $90\%$ exclusion by IceCube-Gen2) are $\nu$-DM interaction, effective operator, significant non-standard neutrino production and Earth matter NSIs. Scenarios with a low exclusion percentage includes standard mixing, quantum decoherence and constant matter effects, however IceCube-Gen2 can still constrain the initial flavor composition considerably in the standard mixing scenario.

It is also potentially interesting to discuss how easy it is to disentangle different scenarios using flavor from the theory perspective only, \ie\, as a matter of principle for an ideal measurement of the flavor composition. Using the same method as before, we quantify this parameter space overlap in \tabl{modeldiscrimination}. We consider ``data'' as the true scenario implemented by Nature, and we ask how much of its parameter space can be discriminated from (lies outside) the ``theory'' scenario. An example: if we believe standard mixing to be correct and ask how much of its parameter space does not coincide with the effective operator scenario, then we find $0$\%  since the standard mixing parameter space lies within the effective operator parameter space. In the opposite situation, one can discriminate the standard mixing scenario in $96$\% of the parameter space, since standard mixing covers only a small fraction of the effective operator parameter space. This means that the table is not symmetric.  Cases with a high discriminating factor are interesting since one can distinguish between them at least in principle. Some examples are: standard mixing vs. $\nu$-DM interaction, constant matter effects vs. decay, and quantum decoherence vs. effective operators.\footnote{Here, we assume the former cases (standard mixing, constant matter effects and quantum decoherence) are test scenarios, and we analyze them against the true cases ($\nu$-DM interaction, decay and effective operator), respectively.} For all the information, we advise the reader to look at \tabl{modeldiscrimination}.

\begin{sidewaysfigure}
\vspace{-0.05\textwidth}
\includegraphics[width=0.45\textwidth]{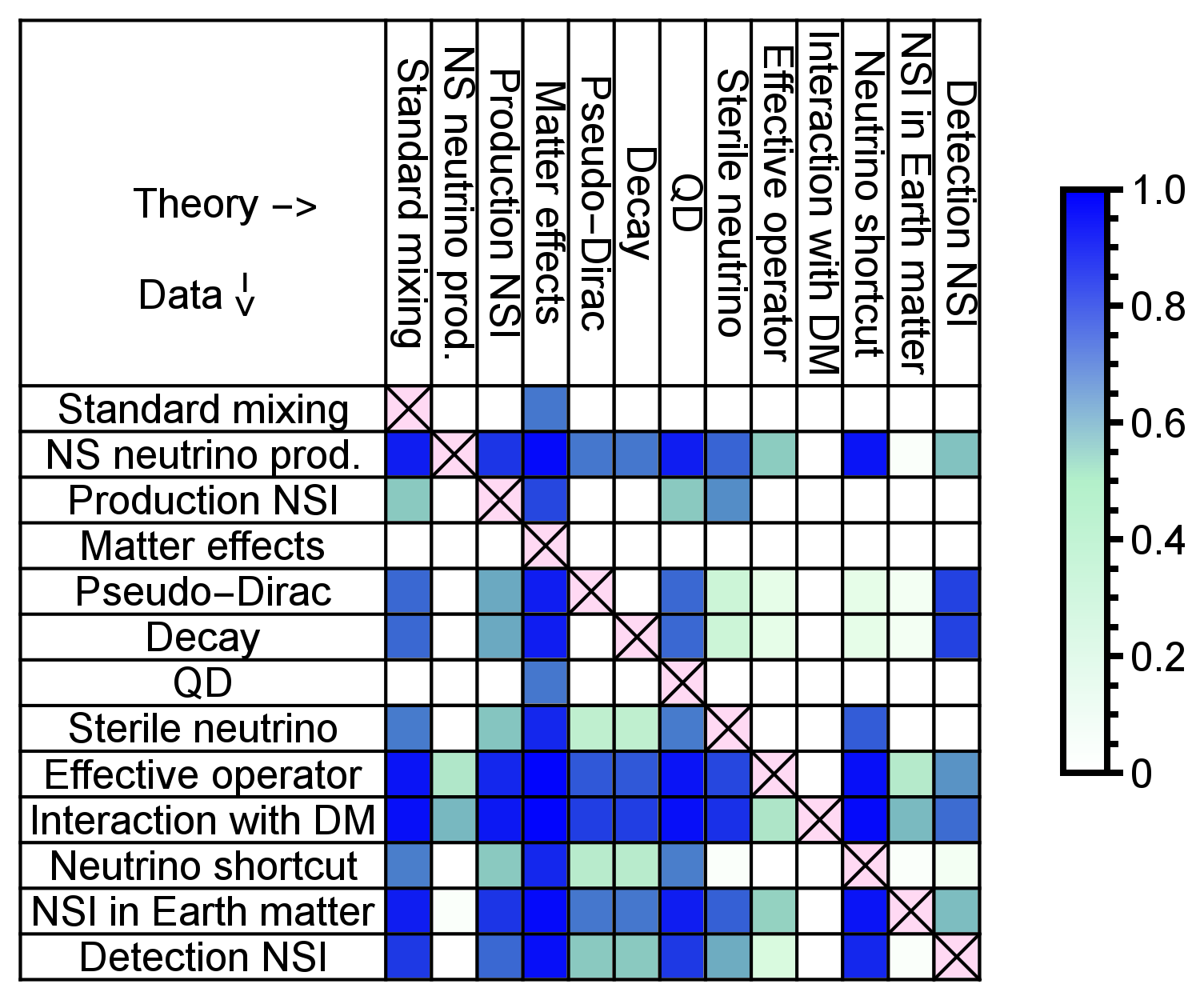}
\caption{Different visualization of \tabl{modeldiscrimination} with ``data'' refering to the scenario implemented by Nature, and ``theory'' to the model to be discriminated. The discrimination percentage is given as a fraction of $100$\%, where a darker (lighter) shading means a higher (lower) discrimination percentage between the scenarios.  Abbreviations: Quantum decoherence (QD), non-standard neutrino production (NS neutrino prod.), Shortcut through extra dimensions (neutrino shortcut).}
\label{fig:modeldiscriminationv2}
\end{sidewaysfigure}

A different visualization of \tabl{modeldiscrimination} is shown in \figu{modeldiscriminationv2} where the discrimination percentage is given as a fraction of $100$~\%. Similar as in \tabl{modeldiscrimination}, the row is the true scenario implemented by Nature (``data''), and the column is the perception of Nature (``theory'' scenario).  A darker (lighter) shading of blue means a higher (lower) discrimination percentage between the scenarios. Take the scenario ``Interaction with DM'' as an example which occupies a large fraction of the parameter space. The other parameter spaces are fully contained within its parameter space. Therefore, one can(not) distinguish between the scenarios when ``Interaction with DM'' is the true (test) case, leading to a dark row and a light column. The scenario ``Matter effects'' is completely opposite to ``Interaction with DM'', since it spans a small fraction of the parameter space, giving it a light row and a dark column in \figu{modeldiscriminationv2}. Half-dark means partially overlapping parameter spaces, meaning one can discriminate about $50$\% of it, independently of the choice of true scenario. This can be compared to the extreme case with complete distinguishable cases or zero discrimination percentage.

\begin{sidewaystable}
\centering
\scriptsize 
\vspace{-0.0\textwidth}
\begin{tabular}{|c|ccccccccccccc|}
\hline
Theory $\rightarrow$ & \multicolumn{1}{l|}{Standard} & \multicolumn{1}{l|}{Non-standard} & \multicolumn{1}{l|}{Production} & \multicolumn{1}{l|}{Matter} & \multicolumn{1}{l|}{Pseudo-Dirac} & \multicolumn{1}{l|}{Decay} & \multicolumn{1}{l|}{Quantum} & \multicolumn{1}{l|}{Sterile} & \multicolumn{1}{l|}{Effective} & \multicolumn{1}{l|}{Interaction} & \multicolumn{1}{l|}{Shortcut througth} & \multicolumn{1}{l|}{NSI in} & \multicolumn{1}{l|}{Detection}   \\ 
Data $\downarrow$ & \multicolumn{1}{l|}{mixing} & \multicolumn{1}{l|}{neutrino production} & \multicolumn{1}{l|}{NSI} & \multicolumn{1}{l|}{effects} & \multicolumn{1}{l|}{neutrino} & \multicolumn{1}{l|}{} & \multicolumn{1}{l|}{decoherence} & \multicolumn{1}{l|}{neutrino} & \multicolumn{1}{l|}{operator} & \multicolumn{1}{l|}{with DM} & \multicolumn{1}{l|}{extra dimension} & \multicolumn{1}{l|}{Earth matter} & \multicolumn{1}{l|}{NSI}   \\ \hline
Standard mixing  &   $\cdots$ & 0\%  &  0\% &  73\%  & 0\% & 0\% & 0\% & 0\% & 0\% & 0\% & 0\% & 0\% & 0\%  \\ \cline{1-1}
Non-standard neutrino production  & 94\% &  $\cdots$ & 89\% &  98\%  & 73\% & 73\% & 94\% & 78\% & 45\% & 0\% & 96\% & 2\% & 49\%   \\ \cline{1-1}
Production NSI &  46\% &  0\% &  $\cdots$ &   85\% & 0\% & 0\% & 46\% & 67\% & 0\% & 0\% & 0\% & 0\% & 0\%  \\ \cline{1-1}
Matter effects &   0\% & 0\% &  0\% & $\cdots$ & 0\%  & 0\% & 0\% & 0\% & 0\% & 0\% & 0\% & 0\% & 0\%    \\ \cline{1-1}
Pseudo-Dirac neutrino &   77\% &  0\% & 58\% &  94\% & $\cdots$  & 0\% & 77\% & 20\% & 10\% & 0\% & 10\% & 5\% & 86\%  \\ \cline{1-1}
Decay &  77\% &  0\% & 58\% &  94\% & 0\% & $\cdots$ & 77\% & 20\% & 10\% & 0\% & 10\% & 5\% & 86\%  \\ \cline{1-1}
Quantum decoherence &  0\% &  0\%  & 0\% & 73\%  & 0\% & 0\% & $\cdots$ & 0\% & 0\% & 0\% & 0\% & 0\% & 0\%  \\ \cline{1-1}
Sterile neutrino &  72\% &  0\%  &  48\% & 92\% & 25\% & 25\% & 72\%  & $\cdots$ & 0\% & 0\% & 80\% & 0\% & 0\%  \\ \cline{1-1}
Effective operator &  96\% &  31\% &  92\% &  99\% & 81\% & 81\% & 96\% & 85\% & $\cdots$ & 0\% & 97\% & 29\% & 65\%  \\ \cline{1-1}
Interaction with DM &  97\% &  53\% &  95\% & 99\% & 87\% & 87\% & 97\% & 90\% & 32\% & $\cdots$ & 98\% & 52\% & 76\%  \\ \cline{1-1}
Shortcut througth extra dimension &  71\% &  0\% & 46\% &  92\% & 28\% & 28\% & 71\% & 2\% & 0\% & 0\% & $\cdots$ & 2\% &  5\%  \\ \cline{1-1}
NSI in Earth matter & 94\% & 2\% & 89\% & 98\% & 73\% & 73\% & 94\% & 79\% & 42\% & 0\% & 96\% & $\cdots$ & 51\% \\ \cline{1-1}
Detection NSI &  88\% &  0\% &  77\% &  97\% & 46\% & 46\% & 88\% & 57\% & 15\% & 0\% & 92\% & 2\% & $\cdots$ \\ \hline
\end{tabular} 
\caption{Quantification of the parameter space overlap between two scenarios.  Here ``data'' refers to the scenario implemented by Nature, and ``theory'' to the model to be discriminated. The numbers give the percentage of the parameter space of ``data'' which can be discriminated from (lies outside) the ``theory'' allowed parameter space in principle (for an ideal measurement).}
\label{tab:modeldiscrimination} 
 \end{sidewaystable}

\section{Discrimination by energy dependence}
\label{sec:energydependence}

\begin{table}[t]
\centering
  \setlength{\tabcolsep}{2pt}
  \begin{tabular}{p{3cm}|p{5cm}}
      \hline
      Scenario  & Input parameters  \\
      \hline 
       Decay &  $\lambda_{2}=\lambda_{3}=100$~s/eV,\par $\text{Br}_{3 \rightarrow 1}=\text{Br}_{3 \rightarrow 2}=0.4$,\par
      $\text{Br}_{3 \rightarrow \text{I}}=0.2$, $\text{Br}_{2 \rightarrow 1}=0.8$, \par
      $\text{Br}_{2 \rightarrow \text{I}}=0.2$, $\lambda_{1}=1000$~s/eV, \par
      $\text{Br}_{1 \rightarrow \text{I}}=1$, $\chi^{2} \leq 11.83$, \par
      $L=100$~Mpc 
      \\
      \hline
      Quantum decoherence &  $\Gamma=3\cdot 10^{-39}$~GeV$^{2}$, \par 
      $\Psi=5\cdot 10^{-39}$~GeV$^{2}$, \par
      $L=100$~Mpc, $\chi^{2} \leq 11.83$, \par
      Chosen energy scaling: \par $\text{e}^{-2\kappa L E^{-1}}$ 
      where $\kappa \in [\Gamma, \Psi]$ 
      \\
      \hline
      Effective operator  &  $O = 9 \cdot 10^{-27}$~GeV, $\Lambda=100$~TeV, \par
      $n=1$ operator $\chi^{2} \leq 11.83$,
      \\
       \hline
      Shortcut through extra dimension & 
      $\theta_{14}=0, \theta_{24}=\theta_{34}=10^{\circ}$, \par $\chi^{2} \leq 11.83$,  $E_{\text{res}}=100$~TeV, \par  $\delta_{24},\delta_{34} \in [0,2\pi]$  
      \\
      \hline     
  \end{tabular}
  \caption{Input parameters used in this section to investigate the energy dependence of the four different scenarios chosen.}
 \label{tab:energydep}
\end{table}

\begin{figure*}[tp]
\centering
\begin{tabular}{c c}
\includegraphics[width=0.35\textwidth]{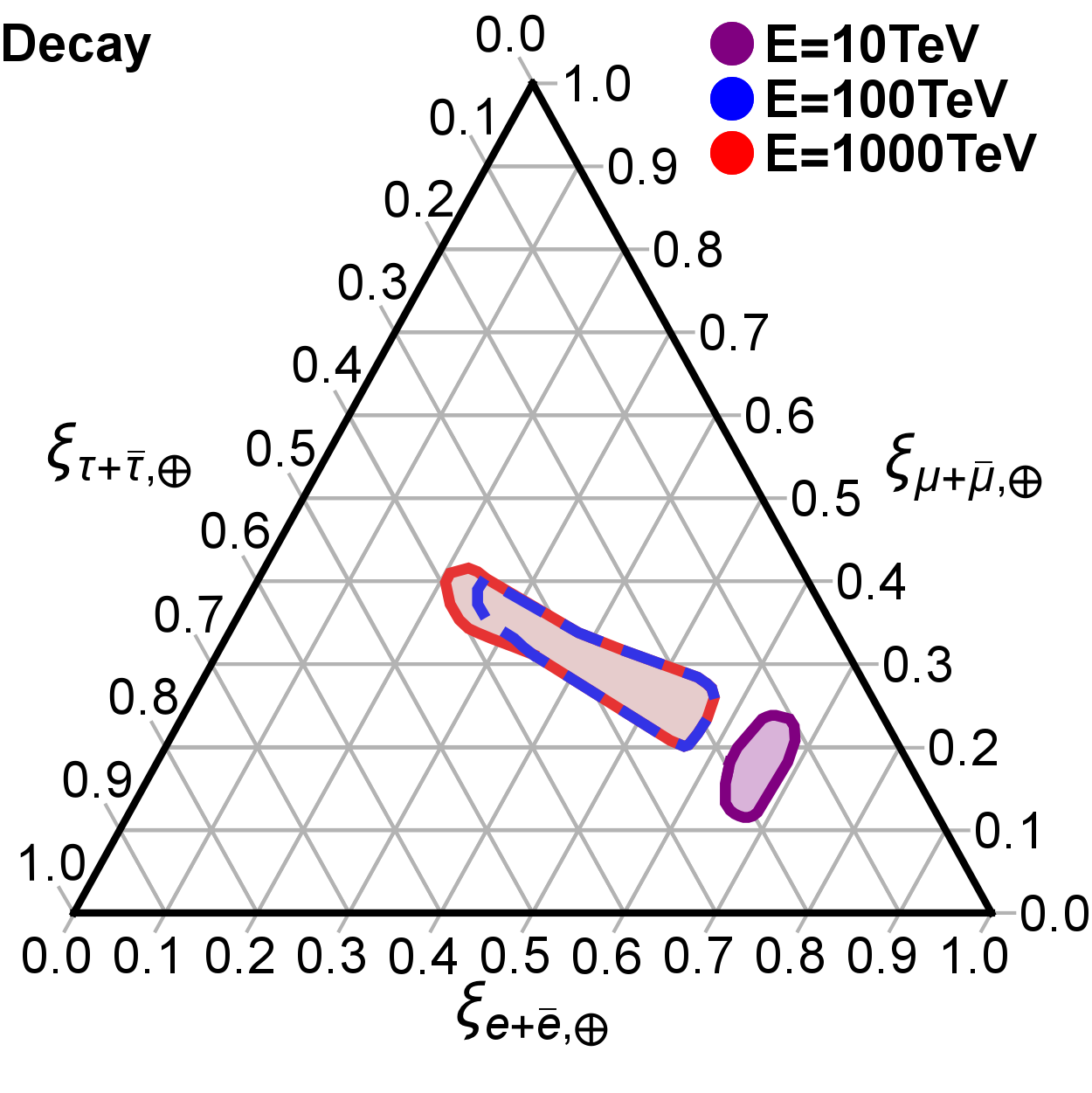}\hspace{0.07\textwidth} &
\includegraphics[width=0.35\textwidth]{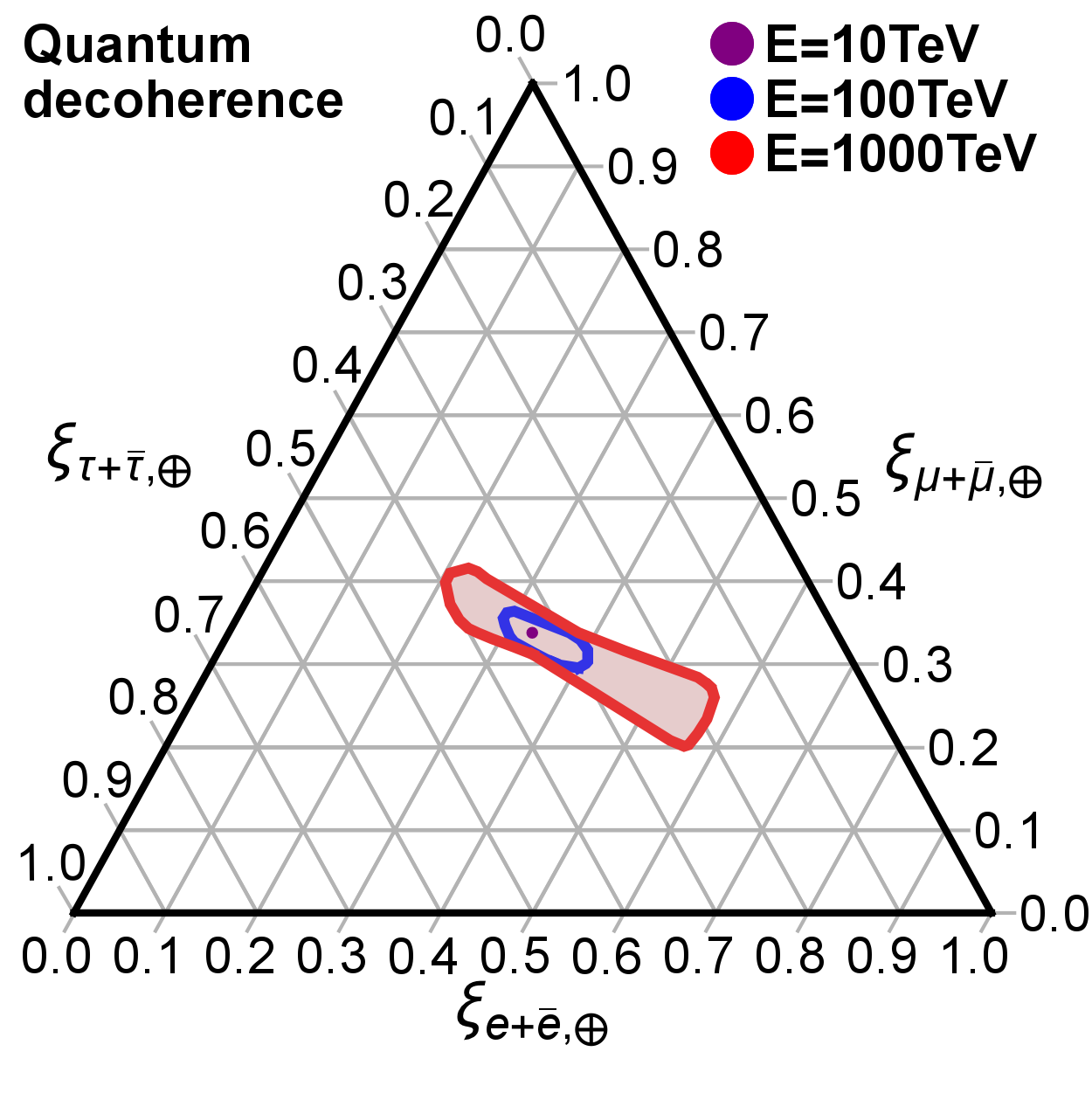} \\
\includegraphics[width=0.35\textwidth]{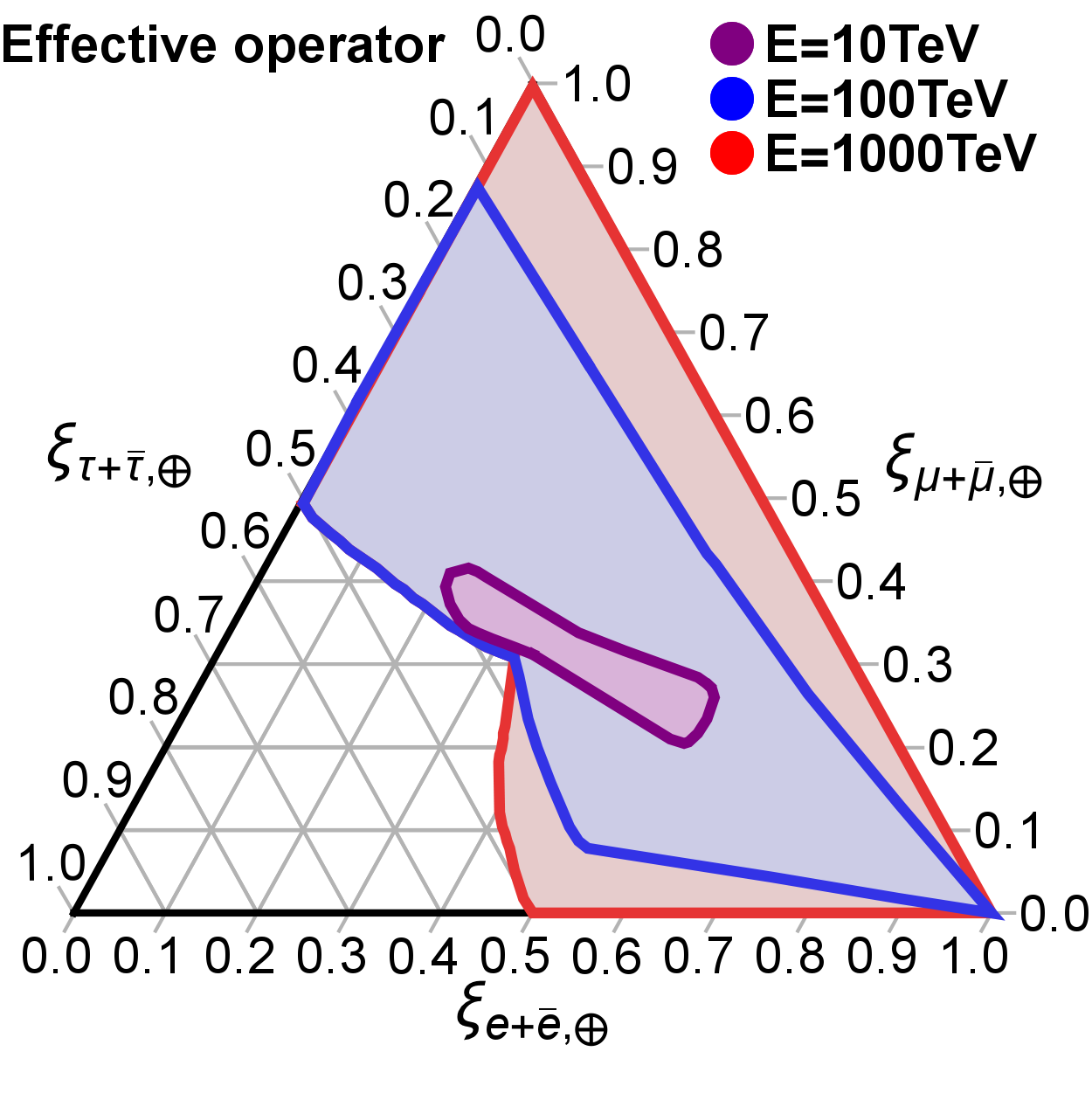} \hspace{0.07\textwidth} &
\includegraphics[width=0.35\textwidth]{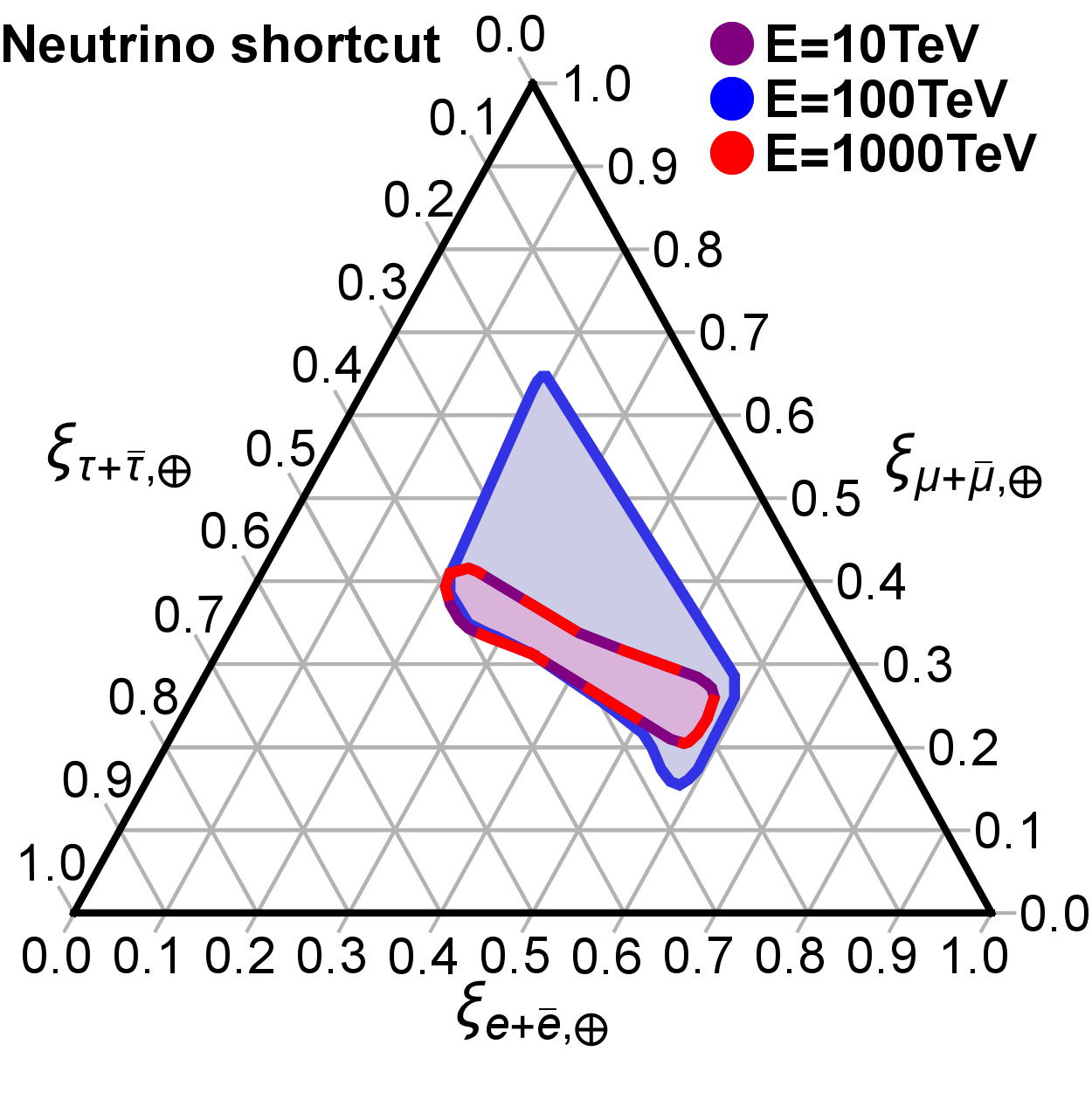} 
\\
\end{tabular}
 \caption{Parameter space for four specific BSM models as a function of energy for the model parameters listed in \Tab~\ref{tab:energydep}. The parameter space of standard mixing coincides with: the red curve for decay, the red curve for quantum decoherence, the purple curve for effective operator and the red/purple curve for sterile neutrino shortcut through extra dimensions. 
}
\label{fig:energydep}
\end{figure*}

So far, we have studied the flavor composition independent of energy (or marginalized over energy). In specific cases, however, the energy dependence can be used to reveal the BSM effect. Here we study three different energies inspired by the potential capability of IceCube-Gen2 \cite{MarekGen2}: 10~TeV, 100~TeV, 1000~TeV. We choose specific scenarios for neutrino decay, quantum decoherence, effective operators and neutrino shortcuts through the extra dimension as examples, see \Tab~\ref{tab:energydep} for the chosen parameter values. We show the result for these four scenarios in \figu{energydep}.

The effect of neutrino decays is typically strongest at low energies (where the Lorentz factor is low), whereas standard mixing is approached for high energies. Quantum decoherence (for the chosen scaling with energy) and effective operators typically show up at high energies, at least for the parameters chosen here. If the energy dependence for quantum decoherence scales such as $\sim e^{-2\kappa L E^{n}}$ with $n=1$, then effects shows up at low energies. The shortcuts through the extra dimension are an example for an effect present in a particular energy range only. Of course, the details (where the transitions occur) depend on the chosen model parameters, but these examples demonstrate that the energy-dependence of the BSM can be used to learn about the BSM physics. For a more detailed discussion of the interplay between a possible energy dependence of the source flavor composition (which we marginalized over here) and energy-dependent BSM physics, see \Ref~\cite{Mehta:2011qb}.

\section{Discrimination by Glashow resonance}
\label{sec:glashowresonance}

\begin{figure}[tp]
\centering
 \includegraphics[width=\columnwidth]{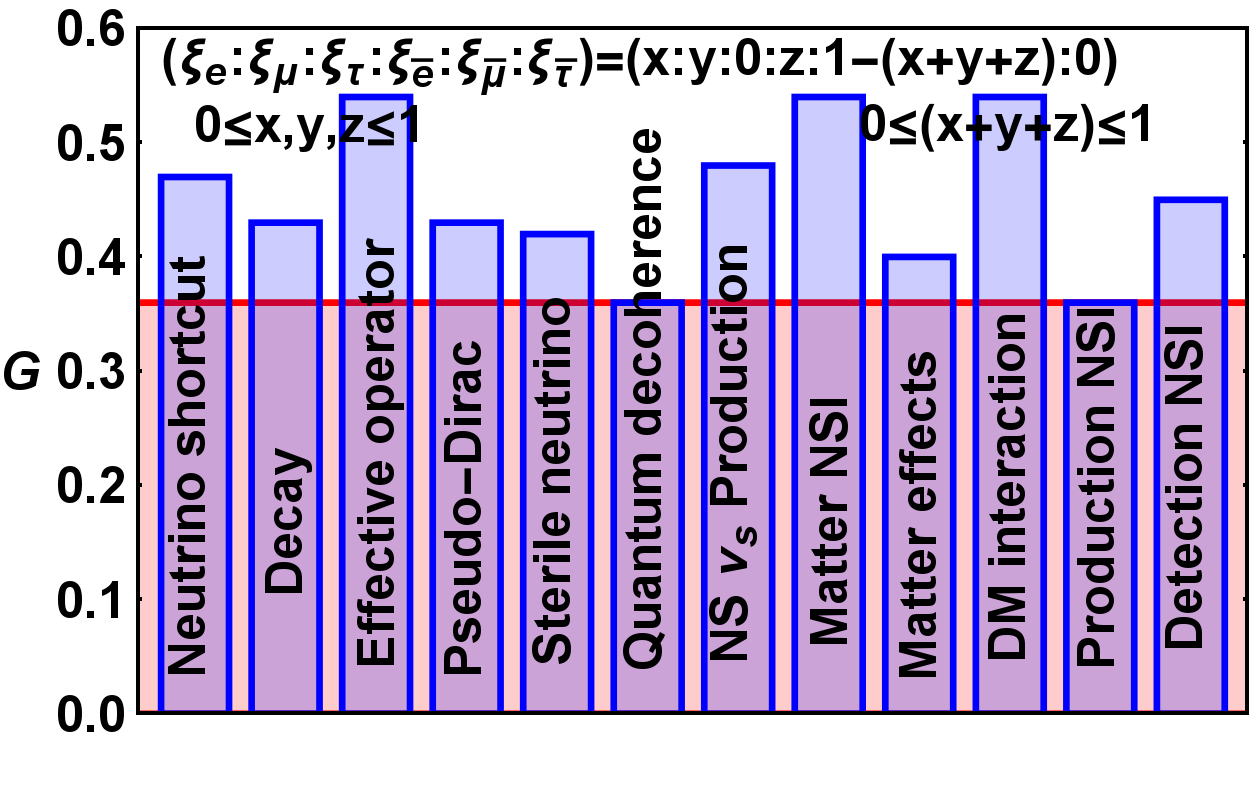}
\caption{Standard mixing (red band) for the electron antineutrino fraction for an arbitrary flavor and neutrino-antineutrino composition at the source, in comparison to the ranges for different BSM scenarios and constant matter effects (bars).}
\label{fig:glashowfreepic}
\end{figure}

Another potential way to distinguish among BSM scenarios is the Glashow resonance, $\bar{\nu}_{e} + e^{-} \rightarrow W^{-} \rightarrow \ \text{anything}$, at $E_{\nu}=m_{W}^{2}/(2 m_{e}) \simeq 6.3$~PeV \cite{PhysRev.118.316}. The Glashow resonance event rate is an indicator for the electron antineutrino contribution to the total flux
\begin{equation}
 G=\frac{\xi_{\bar{e}, \oplus}}{\xi_{e+\bar{e}, \oplus}+\xi_{\mu+\bar{\mu}, \oplus}+\xi_{\tau+\bar{\tau}, \oplus}} \, .
\end{equation}
The Glashow resonance has been used as discriminating power for pp versus p$\gamma$ interactions, which are generic source classes indicative for e.g. starburst galaxies (pp) versus AGNs/GRBs (p$\gamma$)~\cite{Anchordoqui:2004eb}; for a critical discussion see \cite{Biehl:2016psj}.

To obtain the allowed range for $G$, we calculate each BSM scenario and standard mixing with an arbitrary initial electron/muon neutrino flux and neutrino-antineutrino composition in the spirit of the work. We separate the flavor mixing into two channels, one for particles and the other for antiparticles such that we obtain $\xi_{\beta, \oplus}$ and $\xi_{\bar{\epsilon}, \oplus}$ rather than $\xi_{\beta+\bar{\beta}, \oplus}$. One scenario (NSIs in Earth matter) has to be treated with special attention since neutrinos and antineutrinos do not experience the same matter potential, see \Ref~\cite{Gonzalez-Garcia:2016gpq} for further details. To constrain the neutrino oscillation parameters or the mixing matrix elements, we use the appropriate $\chi^{2}$, and we apply the IceCube-Gen2 contours as a constraint. In \figu{glashowfreepic}, the electron antineutrino fraction to the total flux is shown as blue bars for the BSM scenarios and the matter effect case, whereas the overlapping red band is the allowed range for $G$ from standard mixing. Most noticeable, there is no lower bound on $G$ since the neutrino-antineutrino composition at the source is assumed to be unknown. Fixing this quantity, means a lower bound will be present. It is clear that in most cases the additional information from the electron antineutrino fraction will be small compared to the flavor information, at least with the logic applied in this work (unknown flavor and neutrino-antineutrino composition). However, BSM scenarios can predict more Glashow events than standard mixing -- which is a clear signature. 

\section{Diagnostic via direct tau neutrino detection}
\label{sec:tauevent}

\begin{figure}[tp]
\centering
 \includegraphics[width=\columnwidth]{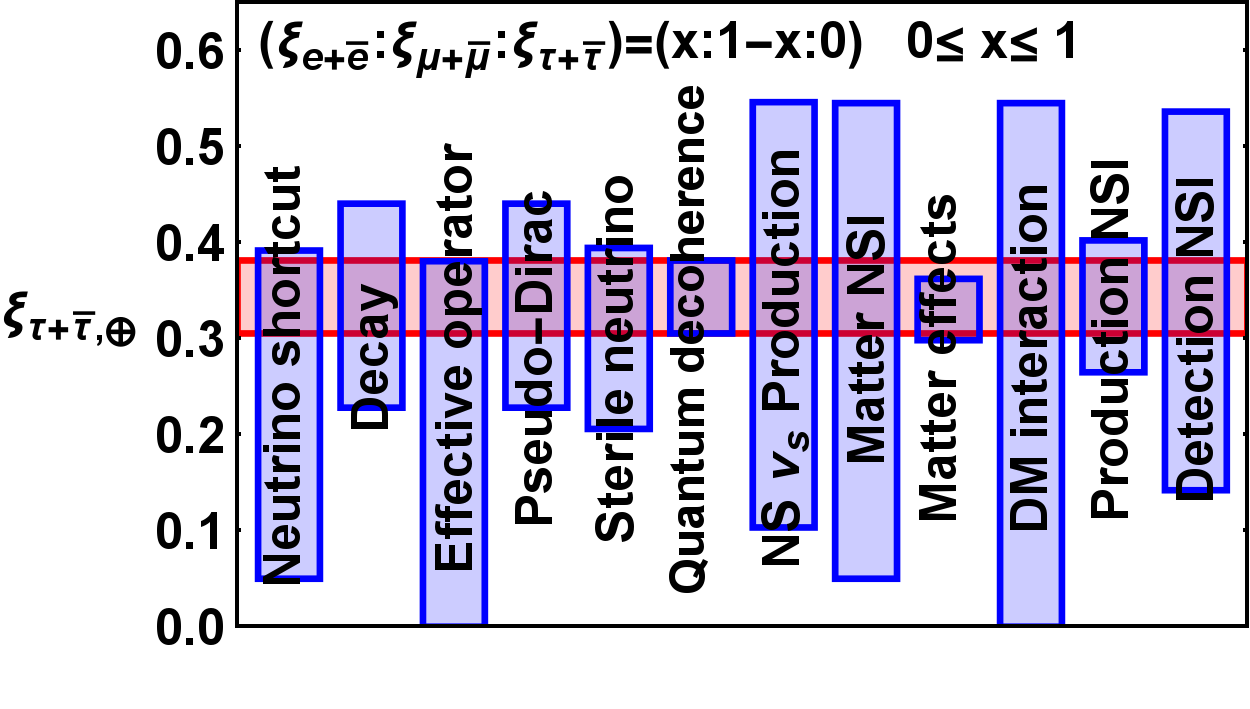}
\caption{The tau flavor composition for standard mixing (red band), in comparison to the ranges for different BSM scenarios and constant matter effects (bars). Large deviations are allowed after the IceCube-Gen2 contours, meaning this additional information can be used to constrain the flavor composition parameter space further.}
\label{fig:tauevent}
\end{figure}

An additional observable to discriminate between new physics scenarios is the number of tau neutrino events, however a signal has not been detected yet. The best known signatures are double bang events \cite{Learned:1994wg}, lollipop events \cite{Beacom:2003nh, DeYoung:2006fg} and double pulse events \cite{Palladino:2015uoa}, which are event topologies one can use to identify tau neutrino events. Other methods \cite{Li:2016kra} can also be used to tag tau neutrinos. Therefore, we present the tau flavor composition in \figu{tauevent}. The red band is the allowed range of the tau flavor composition from standard mixing after applying the IceCube-Gen2 contours as a constraint. The same is done for the BSM scenarios and matter effects, which are represented as blue bars. Standard mixing predicts a small range in comparison to some of the BSM scenarios. 
 
IceCube has recently presented a search for tau neutrinos among the high-energy starting track sample \cite{MarcelICRC2017} with the expectation to identify about 2 tau neutrino events within 6 years of IceCube data. Non were found, nevertheless, we use this signal expectation to scale to IceCube-Gen2. For a contained event sample of $200$~TeV, IceCube-Gen2 will detect about 10 times the event rates of IceCube \cite{MarekGen2}, and hence for 15 years of lifetime, IceCube-Gen2 could see as many as 45 tau neutrino events. The relative error is $15$~\%, meaning a small range of tau neutrino events is expected by standard mixing. One can compute the expected number of tau neutrino events for the BSM scenarios and the constant matter effect scenario by using the tau flavor composition shown in \figu{tauevent}. Comparing this to the range expected from standard mixing, one can constrain the flavor composition parameter space further

\section{Summary and conclusions}
\label{sec:summary}
We have studied the allowed parameter space for the flavor composition at Earth of astrophysical neutrinos above 10~TeV that is allowed by BSM theories. We have used a systematic approach coping with the unknowns such as oscillation parameter uncertainties, the unknown flavor composition at production, and theory model parameters. Our main motivation has been to illustrate the potential of IceCube-Gen2 to study BSM physics by measuring flavor, as astrophysical neutrinos may be one of the best options to search for effect present in extreme environments, at extreme distances, and at extreme energies. We have also included other observables in the discussion such as directional information, the energy-dependence of the flavor composition, Glashow resonance events, and expected tau neutrinos events. 

We have classified the BSM scenarios by production, propagation, and detection effects; \cf, \figu{comparison}. Scenarios with potentially very large deviations from standard mixing include: significant sterile neutrino production (production effect), effective operators from physics at high energy scales or dark matter interactions  (propagation effects), and non-standard interactions in Earth matter (detection effect). We have found that, depending on the BSM scenario, that IceCube-Gen2 can exclude up to 96\% of the allowed parameter space by measuring flavor only.

Further scenario discrimination and parameter identification can be performed using the energy-dependence of the flavor information such as for shortcuts through the extra dimensions, which exhibit strong flavor deviations by a resonance effect, and for effective operators, for which the BSM effect may be naturally expected to kick in at higher energies. The directional (for dark matter interactions) and zenith angle (for non-standard interactions in Earth) can be also used to discriminate among scenarios. While we have demonstrated that the Glashow resonance has limited potential if the flavor and neutrino-antineutrino composition at the source is unknown, whereas tau neutrino events may be an interesting possibility for more precise information on BSM scenarios occupying parameter space closer to the standard mixing expectation.

We conclude that astrophysical neutrinos may be one of the most promising directions to search for BSM physics, complementary to LHC physics, flavor physics, and dark matter searches. While the prime target for IceCube-Gen2 will be searching for the origin of the astrophysics neutrinos, finding physics beyond the Standard Model would be a major breakthrough -- and therefore deserves dedicated experimental and theoretical study.

{\bf Acknowledgments.}  We would like to thank Jakob van Santen, Irene Tamborra, Mauricio Bustamante and Andrea Palladino for valuable discussions and useful suggestions. 

WW has received funding from the European Research Council (ERC) under the European Union’s Horizon 2020 research and innovation programme (Grant No. 646623).

\vspace*{-0.02\textwidth}
\bibliographystyle{apsrev4-1}
\bibliography{references}

\end{document}